\documentclass[10pt]{elsarticle}

\usepackage[utf8]{inputenc}

\usepackage{graphicx}
\usepackage{amsmath}
\usepackage{amssymb}
\usepackage{upgreek}
\usepackage{bm}
\usepackage[version=4]{mhchem}
\usepackage{siunitx}
\usepackage{float}
\usepackage{cleveref}
\usepackage{tikz}
\usetikzlibrary{decorations.pathreplacing}
\usepackage{longtable,tabularx}
\usepackage{indentfirst}
\usepackage{booktabs}
\usepackage{subcaption}
\usepackage{fullpage}

\makeatletter
\def\ps@pprintTitle{%
   \let\@oddhead\@empty
   \let\@evenhead\@empty
   \def\@oddfoot{\reset@font\hfil\thepage\hfil}
   \let\@evenfoot\@oddfoot
}
\makeatother

\begin{document}

\begin{abstract}
A high fidelity multi-physics Eulerian computational framework is presented for the simulation of supersonic
parachute inflation during Mars landing. Unlike previous investigations in this area, the framework takes
into account an initial folding pattern of the parachute, the flow compressibility effect on the fabric
material porosity, and the interactions between supersonic fluid flows and the suspension lines. Several adaptive mesh refinement~(AMR)-enabled, large
edge simulation~(LES)-based, simulations of a full-size disk-gap-band~(DGB) parachute inflating in the
low-density, low-pressure, carbon dioxide (CO$_2$) Martian atmosphere are reported. The comparison of
the drag histories and the first peak forces between the simulation results and experimental data collected
during the NASA Curiosity Rover's Mars atmospheric entry shows reasonable agreements. Furthermore, a
rudimentary material failure analysis is performed to provide an estimate of the safety factor for the
parachute decelerator system.  The proposed framework demonstrates the potential of using Computational
Fluid Dynamics (CFD) and Fluid-Structure Interaction (FSI)-based simulation tools for future supersonic
parachute design.
\end{abstract}

\begin{frontmatter}

\title{Modeling, simulation and validation of supersonic parachute inflation dynamics during Mars landing}

\author[rvt1]{Daniel~Z.~Huang}
\ead{zhengyuh@stanford.edu}

\author[rvt3]{Philip~Avery}
\ead{pavery@stanford.edu}

\author[rvt1,rvt2,rvt3]{Charbel~Farhat}
\ead{cfarhat@stanford.edu}

\author[rvt4]{Jason Rabinovitch}
\ead{jason.rabinovitch@jpl.nasa.gov}

\author[rvt4]{Armen Derkevorkian}
\ead{armen.derkevorkian@jpl.nasa.gov}

\author[rvt4]{Lee D Peterson}
\ead{lee.d.peterson@jpl.nasa.gov}

\address[rvt1]{Institute for Computational and Mathematical Engineering,
               Stanford University, Stanford, CA, 94305}
\address[rvt2]{Mechanical Engineering, Stanford University, Stanford, CA, 94305}
\address[rvt3]{Aeronautics and Astronautics, Stanford University, Stanford, CA, 94305}

\address[rvt4]{Jet Propulsion Laboratory, California Institute of Technology, Pasadena, CA, 91109}

\end{frontmatter}

\section*{Nomenclature}
{\renewcommand\arraystretch{1.0}
\noindent\begin{longtable*}{@{}l @{\quad=\quad} l@{}}
{$H_{\Box}$}               &{parachute system length parameter}\\
{$D_{\Box}$}               &{parachute system diameter parameter}\\
{$\Omega^{\Box}$}          &{computational domain}\\
{$\partial \Omega^{\Box}$} &{boundary of the computational domain}\\
{$\Omega^F_{e}$}           &{fluid mesh primal cell}\\
{$\mathcal{C}_{i}$}        &{fluid mesh dual cell or control volume}\\
{$\mathcal{\partial C}_{i}$} &{fluid mesh dual cell boundary}\\
{$\bm{\nu}_{ij}$}          &{area weighted dual cell facet normal}\\
{$N_n^e$}                  &{number of nodes attached to the primal cell $\Omega^F_{e}$}\\
{$\phi_i$}                 &{shape function associated with node $i$}\\
{$\mathcal{F}$}            &{fluid inviscid flux tensor}\\
{$\bm{\Phi}_{ij}$}         &{approximate Riemann flux}\\
{$\mathcal{G}$}            &{fluid viscous flux tensor}\\
{$\bm{W}$}                 &{fluid conservative variable}\\
{$\bm{V}$}                 &{fluid primitive variable}\\
{$\bm{v}$}                 &{fluid velocity}\\
{$v_{i}$}                  &{fluid velocity component}\\
{$\rho^F$}                 &{fluid density}\\
{$p$}                      &{pressure}\\
{$\mathcal{E}$}            &{total energy per unit volume}\\
{$e$}                      &{specific internal energy}\\
{$T$}                      &{temperature}\\
{${\uptau}$}               &{viscous stress tensor}\\
{$\bm{q}$}                 &{heat flux vector}\\
{$\mu_{\Box}$}             &{viscosity}\\
{$\mathsf{Ma}$}            &{Mach number}\\
{$R$}                      &{gas constant}\\
{$\gamma$}                 &{specific heat ratio}\\
{$\kappa$}                 &{thermal conductivity}\\
{$\gamma$}                 &{specific heat at constant pressure}\\
{$\mathsf{Pr}$}            &{Prandtl number}\\
{$\mathcal{I}$}            &{identity matrix}\\
{$\rho^S$}                 &{structure density}\\
{$\mathbb{M}$}             &{mass matrix} \\
{$E$}                      &{Young's modulus}\\
{$\nu$}                    &{Poisson's ratio}\\
{$th$}                     &{fabric thickness}\\
{$\bm{u}$}                 &{structure displacement}\\
{$u_{i}$}                  &{structure displacement component}\\
{$\bm{\theta}$}            &{structure rotation vector}\\
{$\omega$}                 &{structure angular velocity}\\
{$\mathcal{R}$}            &{rotation matrix}\\
{$\bm{f}$}                 &{force vector}\\
{$\bm{m}$}                 &{moment vector}\\
{$\bm{b}$}                 &{body force}\\
{$\bm{E}$}                 &{Green strain tensor}\\
{$\bm{F}$}                 &{deformation gradient tensor}\\
{$J$}                      &{determinant of the deformation gradient tensor $\bm{F}$}\\
{$\bm{S}$}                 &{second Piola-Kirchhoff stress tensor}\\
{$\Sigma$}                 &{fluid-structure interface}\\
{$\bm{n}$}                 &{normal of the fluid-structure interface}\\
{$\alpha$}                 &{void fraction or porosity}\\
{$\bm{x}$}                 &{Eulerian coordinates}\\
{$\bm{d}$}                 &{distance vector}\\
{$\bm{X}$}                 &{material coordinates}\\
{$S_{i}^{j}$}              &{slave node}\\
{$M_{i}$}                  &{master point}\\
{$n_i$}                    &{number of slave nodes paired with the master point $M_i$} \\
{$t$}                      &{time}\\

\multicolumn{2}{@{}l}{Subscripts}\\
{$\Box_{\infty}$}         &{far-field related quantities}\\
{$\Box_{h}$}              &{discretized state}\\

\multicolumn{2}{@{}l}{Superscripts}\\
{$\Box^{S}$}              &{structure related quantities}\\
{$\Box^{F}$}              &{fluid related quantities}\\

\multicolumn{2}{@{}l}{Operators}\\
{$\Box^{T}$}              &{transpose operator}\\
{$\dot{\Box}$}            &{time derivative}\\
{$\nabla \Box$}           &{gradient operator}\\
\end{longtable*}}

\section{Introduction}
Parachute decelerator systems have been used to successfully land large mass spacecraft on the surface
of Mars since the 1970s \cite{cooley1977viking, spencer1999mars, prakash2008mars, witkowski2003mars,
adams2011phoenix}. However, the dynamics of parachute inflation at supersonic speeds involves complex
interdependent phenomena such as interactions between shocks, turbulent wakes and flexible membrane deformations,
the ramifications of which still remain unclear.  These complex systems have been studied experimentally
(mostly in the 1960s) by the US Air Force and subsequently by NASA during the qualification of the Viking
program. It has been reported that the flow conditions prevalent in the supersonic flow regime have a
profound effect on the performance of a parachute. In particular a high-frequency, large-amplitude oscillation
know as breathing was observed for Mach numbers between 1.6 and 4.65 \cite{maynard1961aerodynamic, charczenko1964wind}.
Furthermore, a parachute may suffer from a rapid fall-off of drag performance at supersonic speeds
\cite{babish1966drag, sengupta2011fluid}, despite performing well in subsonic air-streams. Cause for even
greater concern are the the numerous canopy failures observed in flight tests conducted over the past
few decades, which typically occurred during inflation \cite{o2016reconstructed, rabinovitch2018preliminary}, 
and the limited number of tests that can be performed for developing a better understanding of how to
avoid such failures. Despite this, most if not all relevant computational efforts have focused on developing
CFD and FSI parachute models for the \emph{post-inflation} regime \cite{karagiozis2011computational,
lingard2005simulation, barnhardt2007detached, sengupta2008supersonic}, which is substantially easier
to simulate; few predictive high fidelity simulations of the dynamics of a parachute during its inflation
have been reported \cite{gao2016numerical, huang2018simulation}. Consequently, the Jet Propulsion Laboratory,
California Institute of Technology, and Stanford University have initiated a multiyear research collaboration
with the goal of advancing the state-of-the-art for modeling the supersonic parachute inflation process.
This paper presents results from a numerical study that aims to understand the underlying physics and
emphasizes the validation with Mars landing supersonic parachute data.

The parachute inflation process for typical NASA Mars landed missions starts with a mortar fire \cite{cruz2014reconstruction}, during which
the parachute pack is ejected by the mortar deployment system at a specific ejection velocity to enable
successful bag strip and suspension line stretch without line tangling or damage to the parachute during deployment.
After suspension line stretch, the parachute inflates rapidly to an initial peak force, followed by one
or more partial collapse cycles \cite{cruz2014reconstruction}. Simulation of the first part of this process
from the mortar fire until suspension line stretch presents numerous challenges and is beyond the scope
of the present study. Consequently, the simulation in the present work focuses on the second part of
inflation process commencing at the line stretch stage, with an initial flow condition characterized
by turbulence and shock waves carefully established to mitigate any artificial effects. The influence
of parachute folding -- which is generally ignored in the literature -- on the inflation process and
initial stress of the parachute fabric was reported in \cite{huang2018simulation} to be significant.
Hence, an idealized analytical representation of the line stretch configuration based on a partial unfolding
with associated prestress is used for the structure initial state of the parachute in this work.

During inflation, the parachute interacts with supersonic flows and undergoes large deformation and possibly
even topological changes. This makes arbitrary Eulerian Lagrangian (ALE) methods \cite{hirt1974arbitrary,
donea1982arbitrary} unfeasible, even with remeshing techniques \cite{tezduyar2007modelling}. Hence, the
Eulerian computational framework with an immersed (embedded) boundary method \cite{peskin1972flow, mittal2005immersed}
is adopted in the present study. Specifically, the Finite Volume method with Exact two-phase or two-material
Riemann problems (FIVER) \cite{farhat2008higher, wang2011algorithms, lakshminarayan2014embedded, huang2018family}
is utilized. This method has previously been successfully employed for the simulation of the failure analysis
of submerged structures subjected to explosions and implosions \cite{farhat2013dynamic, wang2015computational}.
It incorporates in the framework a parallel adaptive mesh refinement based on newest vertex bisection
\cite{mitchell1988unified,stevenson2008completion,borker2019mesh}, which enables the capturing of various
interactions between the fluid subsystem, the nonlinear parachute subsystem, and the forebody. It also
enables resolution of all relevant self-contact \cite{heinstein2004acme} effects of the parachute during
its inflation. This paper overviews several novel techniques recently developed specifically for parachute
inflation simulations in this computational framework. They include a homogenized porous wall model that
takes into account the flow compressibility, and a suspension line treatment that allows for the interactions
of the sub-grid scale suspension lines with the flow. These ingredients have been reported to play a
major role in parachute drag performance and overall stability, based on wind-tunnel testing \cite{heinrich1966aerodynamics,
sengupta2008supersonic, cruz2003wind}, but have been generally neglected or oversimplified by empirical
models from the literature \cite{kim20062, gao2016numerical, gao2016bnumerical, tezduyar2007modelling}.
 
The validation of the framework is presented by comparing simulation results with data collected by the
NASA Curiosity Rover during its Mars atmospheric entry \cite{cruz2014reconstruction}. Namely, a full-scale
parachute inflated in the low-density, low-pressure Mars atmosphere is simulated by utilizing Adaptive
Mesh Refinement (AMR) and Large-Eddy Simulation (LES) of compressible flow, coupled with a highly flexible
structure. The canopy breathing, bow shocks and turbulent wake interactions are observed. Comparisons
of the drag history and the first peak force with the collected data are presented; reasonable agreements
are achieved, even though the forebody is assumed to be rigid without vibrations/oscillations, and the trim angle is set
to zero. To the best of our knowledge, this is the first FSI simulation that reasonably matches the Mars landing data.
 
The aim of the present study is two-fold. First, to overview the techniques developed for parachute inflation
simulations. Second, to demonstrate the ability of state-of-art CFD and FSI techniques to accurately predict
the first peak and the maximum von Mises stress of supersonic parachute inflation at a manageable computational
cost. The remainder of this paper is organized as follows. \Cref{sec:setup} introduces the setup of the
simulation including the choice of the suitable initial conditions. \Cref{sec:numerical} briefly outlines
the mathematical formulation and computational models underlying the adopted dynamic fluid and structural
models and their interactions. \Cref{sec:validation} compares critically the results from simulations with
in-flight measurements. Finally, conclusions are offered in~\cref{sec:conclusion}.

\section{Problem setup}
\label{sec:setup}

\begin{figure}
\centering
\includegraphics[scale=0.40]{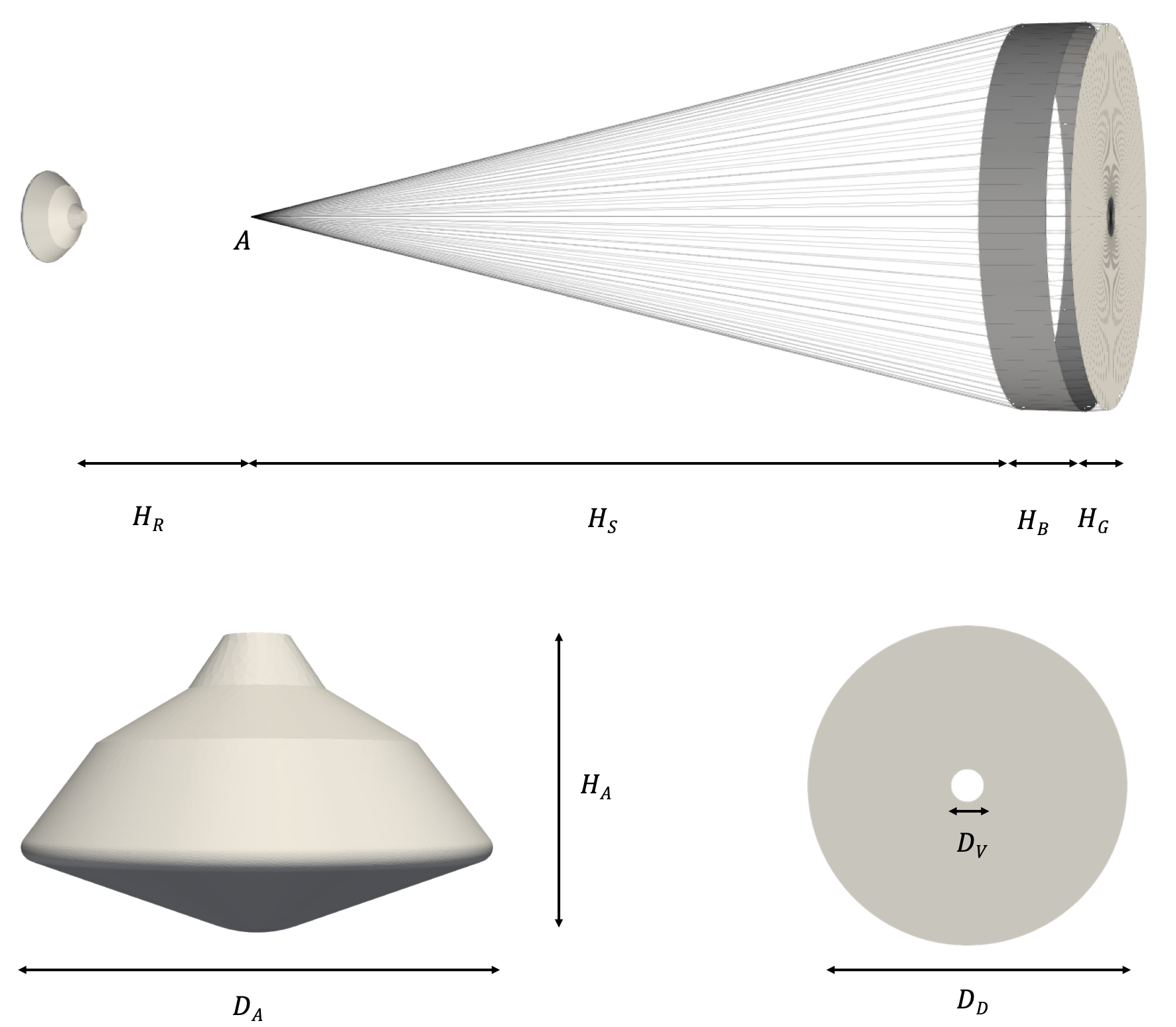}
\caption{Geometry of the disk-gap-band parachute system including the MSL-like entry vehicle.}
\label{fig:parachute_geo}
\end{figure}

The parachute decelerator system considered in the present study is shown in~\cref{fig:parachute_geo}.
It is that which successfully landed the Curiosity Rover on the surface of Mars in 2012. The system
consists of three main components:
\begin{itemize}
\item the DGB parachute canopy comprising a disk part and a band part separated by gap for stability
      purposes, each primarily made of nylon fabric,
\item the suspension lines, which are made of Technora cord (also referred to in the paper as the cords), and
\item the Mars Science Laboratory~(MSL)-like entry vehicle.
\end{itemize}

Each of the 40 cords form a continuous loop starting from the confluence point $A$, then passing through
the band, disk~(via the central vent) and band once again (on the opposite side), before finally terminating
back at $A$. This arrangement therefore essentially corresponds to a configuration with 80 suspension lines.
The cables are sewed to the fabric of the canopy along the regions in which they align. Additionally,
four cables are integrated as radial load carrying members in the band and disk leading and trailing edges. 
The DGB parachute canopy consists of 80 gores, with each gore comprising an approximately triangular segment
in the disk and a rectangular segment in the band. Detailed geometric parameters \footnote{The dimensions
of the MSL-like entry vehicle are assumed to be similar to those published in \cite{clark2009supersonic}.}
\cite{cruz2014reconstruction, clark2009supersonic} and material properties~\cite{lin2010flexible} are listed
in \cref{tab:parachute_system}.

\begin{table}[!htbp]
\centering
\begin{tabular}{@{}lll@{}}
\toprule
 Parameter   & Description                            & Value                      \\
\midrule
 $H_{R}$     & Riser length~(including triple bridle) & 8.895 m                    \\
 $H_{S}$     & Suspension line height                 & 35.826 m                   \\
 $H_{B}$     & Band height                            & 2.580 m                    \\
 $H_{G}$     & Gap height                             & 0.904 m                    \\
 $H_A$       & Entry vehicle height                   & 2.7647m                    \\
 $D_A$       & Entry vehicle diameter                 & 4.5 m                      \\
 $D_V$       & Vent diameter                          & 1.576 m                    \\
 $D_D$       & Disk~(band) diameter                   & 15.447m                    \\
 $D_S$       & Suspension line diameter               & 3.175 $\times$ 10$^{-3}$ m \\
\midrule 
 $E$         & Fabric Young's modulus                 & 9.448$\times$10$^8$ Pa     \\
 $\nu$       & Fabric Poisson's ratio                 & 0.4 \\
 $th$        & Fabric thickness                       & 7.6073$\times$ 10$^{-5}$ m \\
 $\rho^S$    & Fabric density                         & 1154.25 kg m$^{-3}$        \\
 $E'$        & Suspension line Young's modulus        & 2.951$\times$10$^{10}$ Pa  \\
 $\nu'$      & Suspension line Poisson's ratio        & 0.4                        \\
 $\rho'^{S}$ & Suspension line density                & 1154.25 kg m$^{-3}$        \\
\bottomrule
\end{tabular}
\caption{Geometric parameters and material properties of the parachute decelerator system.}
\label{tab:parachute_system}
\end{table}

\begin{figure}
\centering
\includegraphics[scale=0.5]{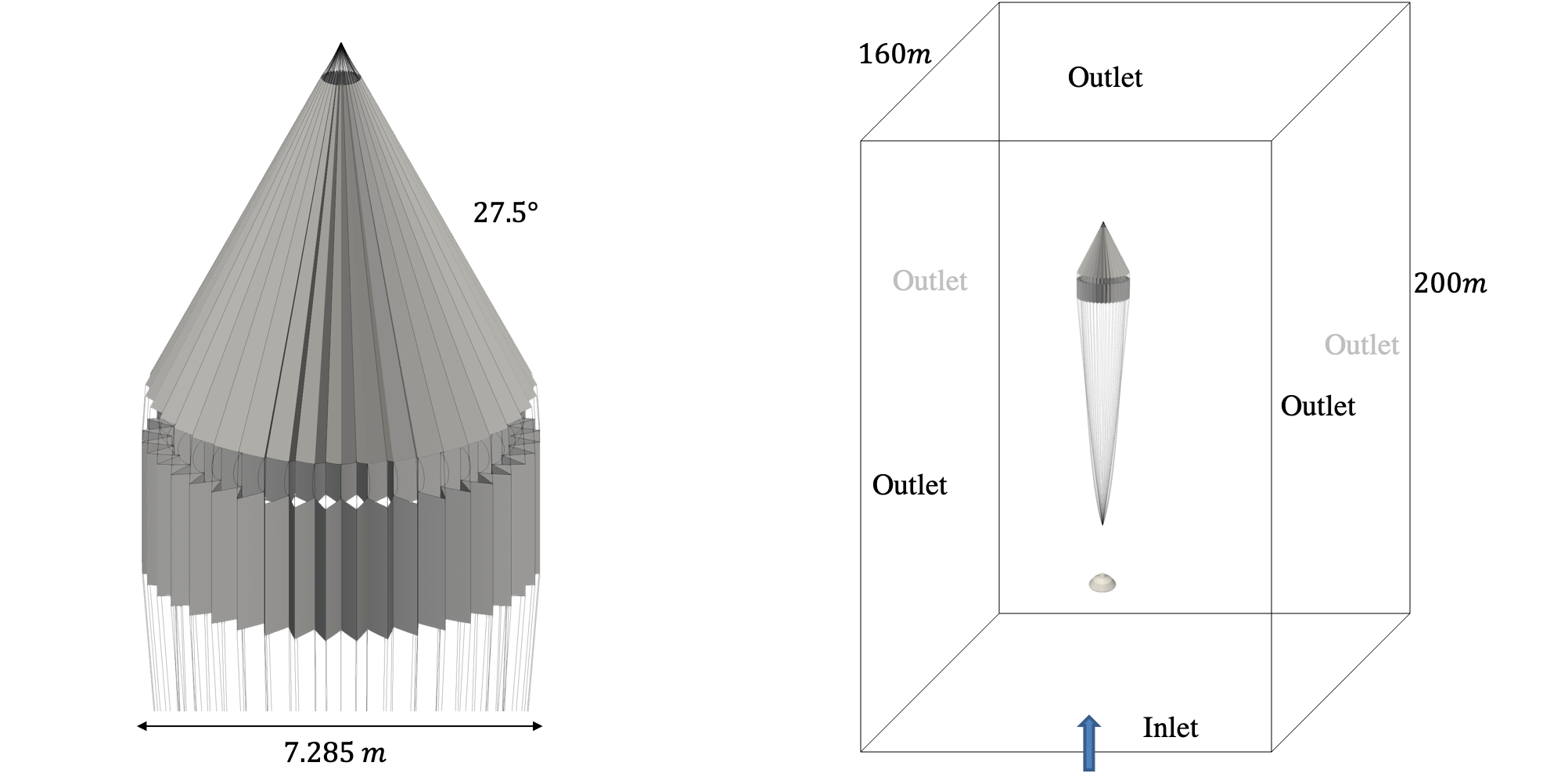}
\caption{Problem setup of a supersonic parachute inflation dynamics: geometry of the initial parachute
         folding~(left) and computational fluid domain~(right).}
\label{fig:parachute_initial}
\end{figure}

When the spacecraft reaches the Martian atmosphere, it decelerates due to the drag of the entry vehicle, until the velocity reaches below Mach 2. Then, the parachute pack is ejected by a
mortar deployment system at a specific ejection velocity. Simulation of the initial deployment and unpacking
of the parachute pack commencing from the mortar fire stage presents numerous challenges that are beyond
the scope of the present work. Instead, the simulation in the present work is started from the line stretch
stage, with an initial flow condition characterized by turbulence and shock waves carefully established
to mitigate any artificial effects, and an idealized analytical representation of the line stretch configuration
for the parachute canopy (see~\cref{fig:parachute_initial}) constructed by rigidly rotating each gore from the
as-built configuration which is assumed to be stress-free. The folding angles are $23.5^{\circ}$ and
$27.5^{\circ}$ for inner and outer folds of the disk~(from the center line to the folds), respectively,
resulting in a projected diameter of $d=7.285$~m. The suspension lines may be specified as either straight
lines or as catenary curves to ensure full length. The corresponding pre-stress due to the folding is
also taken into account. 

To obtain the initial flow conditions, a sequence of fluid simulations were undertaken. First, a quasi
steady-state solution was computed for the flow past the fixed entry vehicle by simulating the evolution
of the flow starting from a uniform initial (free-stream) condition and terminating after $0.15$~s elapsed
time, without including the parachute canopy or suspension lines. Then the simulation is restarted and
pursued for a further $0.15$~s with the the parachute canopy and suspension lines included but fixed
in their analytical line stretch configuration as shown in~\cref{fig:parachute_initial}. During these
two simulations, bow shocks form in front of both the canopy and the entry vehicle. Finally, the coupled
FSI problem associated with the inflation of the DGB parachute starting from the previously computed
fluid state is simulated in the time interval [0 s, 0.8 s], during which the inflation process is expected
to complete and several breathing cycles are expected to be captured. In all three simulations, the spatial
computational domain of the fluid is defined as [-80 m, 80 m] $\times$ [-80 m, 80 m] $\times$ [-20 m, 180 m]
with the confluence point $A$ centered at the origin (see~\cref{fig:parachute_initial}).

\section{Computational modeling}
\label{sec:numerical}

\subsection{Governing equations}
\subsubsection{Governing equations of the fluid subsystem}
The conservation form of the Navier-Stokes equations governing the motion of the fluid can be written as
\begin{equation}
\label{eq: ns}
  \frac{\partial \bm{W}}{\partial t} + \nabla\cdot \mathcal{F}\left(\bm{W}\right) = \nabla\cdot \mathcal{G}\left(\bm{V},
  \nabla\bm{V}\right) \quad \mathrm{in} \; \Omega^F,
\end{equation}
where $\bm{V}$ and $\bm{W}$ are the vectors of the primitive and conservative variables describing the
fluid state, $\mathcal{F}(\bm{W})$ and $\mathcal{G}(\bm{V}, \nabla\bm{V})$ are the inviscid
and viscous flux tensors, and $\Omega^F$ denotes the fixed computational fluid domain. Specifically,
\begin{equation*}
  \bm{V} = \begin{pmatrix} \rho^{F}   \\[4mm]
                           \bm{v}     \\[4mm]
                           p
           \end{pmatrix}, \quad
  \bm{W} = \begin{pmatrix} \rho^{F}        \\[4mm]
                           \rho^{F} \bm{v} \\[4mm]
                           \mathcal{E}
           \end{pmatrix}, \quad 
  \mathcal{F}(\bm{W}) = \begin{pmatrix} \rho^{F} \bm{v}                               \\[2mm]
                                        \rho^{F} \bm{v} \otimes \bm{v} + p\mathcal{I} \\[2.5mm]
                                        \bigl(\mathcal{E} + p \bigr)\bm{v}
                        \end{pmatrix}, \quad \mathrm{and} \quad
  \mathcal{G}(\bm{V},\nabla\bm{V}) = \begin{pmatrix} 0          \\[2.5mm]
                                                    {\bm{\tau}} \\[1.5mm]
                                                    {\bm{\tau}}\boldsymbol \cdot \bm{v} - \bm{q}
                                     \end{pmatrix},
\end{equation*}
where $\rho^{F}$, $\bm{v}$, $p$, and $\mathcal{E}$ denote the density, velocity vector, static pressure, and total energy
per unit volume of the fluid, respectively. The velocity vector and total energy per unit volume are given by
\begin{equation*}
  \bm{v} = \left(v_1, v_2, v_3\right)^{T} \quad \textrm{and} \quad \mathcal{E} = \rho^{F} e + \frac{1}{2}\rho^{F} \left(v_1^2 + v_2^2 + v_3^2\right),
\end{equation*}
where $e$ denotes the specific (i.e., per unit of mass) internal energy. In the definition of the inviscid
flux tensor $\mathcal{F}$, $\mathcal{I} \in \mathbb{R}^{3 \times 3}$ denotes the identity matrix. In
the definition of the viscous flux tensor $\mathcal{G}$, $\bm{\tau}$ and $\bm{q}$ denote the viscous
stress tensor and heat flux vector, respectively, and are defined as
\begin{equation*}
  \bm{\tau} = \mu\left( \nabla^T \bm{v} + \nabla \bm{v}\right) + \left(\mu_v - \frac{2}{3}\mu\right)\left(\nabla\boldsymbol{\cdot}
              \bm{v}\right)\mathcal{I}, \quad \mathrm{and} \quad
  \bm{q} = -\kappa \nabla T,
\end{equation*}
where $\mu$ is the dynamic viscosity, $\mu_v$ is the bulk viscosity, $\kappa$ is the thermal conductivity,
and $T$ is the temperature. The dynamic viscosity is modeled using Sutherland's viscosity law
\begin{equation*}
  \mu = \frac{\mu_0 \sqrt{T}}{1 + T_0/T}
\end{equation*}
where $T_0$ is the reference temperature and $\mu_0$ is the corresponding viscosity. The thermal conductivity
is defined, given a constant Prandtl number ($\mathsf{Pr}$), as
\begin{equation*}
  \kappa = \frac{c_p \mu}{\mathsf{Pr}}
\end{equation*}
where $c_p$ is the specific heat at constant pressure.

The system of equations~(\ref{eq: ns}) is closed by assuming that the gas is ideal and calorically perfect
\begin{equation*}
  p = \rho^{F} R T \quad \mathrm{and} \quad e = \frac{R}{\gamma - 1} T,
\end{equation*}
where $R$ and $\gamma$ are the gas constant and specific heat ratio, respectively. The specific heat
at constant pressure is given by 
\begin{equation*}
  c_p = \frac{R\gamma}{\gamma-1}.
\end{equation*}
The flow is assumed to have transitioned to the turbulent regime, which is modeled using the Vreman turbulence
model~\cite{vreman2004eddy}. The atmosphere of Mars is primarily composed of CO$_2$, and its parameters
are listed in \cref{tab:co2}. It is worth mentioning that the bulk viscosity of CO$_2$ has been reported to be significantly
larger than its dynamic viscosity \footnote{The implementation of a large constant (compared to temperature or frequency dependent, for example) bulk viscosity of CO$_2$ in the Navier-Stokes equation
is still under investigation the exploration of this is out of the scope of present work. Simulation results
both with and without considering bulk viscosity are presented.}; in the present study it is set to be 1000 times
greater than the dynamic viscosity ~\cite{emanuel1990bulk,emmanuel1992bulk,graves1999bulk,jaeger2018bulk,meador1996bulk}.

\begin{table}[!htbp]
\centering
\begin{tabular}{@{}lll@{}}
\toprule
 Parameter         & Description                                         & Value                                      \\
\midrule
 $\mu_0$           & Reference viscosity in Sutherland's viscosity law   & 1.57$\times$10$^{-6}$ kg m$^{-1}$ s$^{-1}$ \\
 $T_0$             & Reference temperature in Sutherland's viscosity law & 240 K                                      \\
 $\mu_v/\mu$       & Bulk viscosity ratio                                & 1000                                       \\
 $R$               & Gas constant                                        & 188.4 J mol$^{-1}$ K$^{-1}$                \\
 $\gamma$          & Specific heat ratio                                 & 1.33                                       \\
 $\mathsf{Pr}$     & Prandtl number                                      & 0.72                                       \\
 $\mathsf{{Pr}_t}$ & Turbulent Prandtl number                        & 0.9                                        \\
 $C_s$             & Constant used in Vreman's turbulence model          & 0.07                                       \\
\bottomrule
\end{tabular}
\caption{Parameters of the Martian atmosphere~(CO$_2$) used in the present study.}
\label{tab:co2}
\end{table}

\subsubsection{Governing equations of the structural subsystem}
The governing equations of the dynamic equilibrium of the structure (including the cable subsystem) are
written in the Lagrangian formulation
\begin{equation}
\label{eq: struct_equations}
  \rho^S \ddot{\bm{u}} - \nabla_{\bm{X}} (\bm{F}\bm{S}) = \bm{b} \quad \mathrm{in} \; \Omega_0^S,
\end{equation}
where $\Omega_0^S$ denotes the initial configuration with material coordinates $\bm{X}$, $\rho^S$ denotes
the structural material density, $\bm{u}=(u_1,u_2,u_3)$ denotes the displacement vector, $\bm{F}$ denotes
the deformation gradient tensor, $\bm{S}$ denotes the symmetric second Piola-Kirchhoff stress tensor, and $\bm{b}$
is the vector of body forces acting on $\Omega_0^S$. Given a structural material of interest, the closure
of~\cref{eq: struct_equations} is performed by specifying a constitutive relation that typically relates
the second Piola-Kirchhoff stress tensor to the symmetric Green strain tensor ($\bm{E}$), defined as
\begin{equation*}
  \bm{E} = \frac{1}{2}\left(\bm{F}^T \bm{F} - \mathcal{I}\right).
\end{equation*}
Dirichlet and/or Neumann boundary conditions are applied to the Dirichlet and Neumann part of the boundary
of $\Omega_0^S$, as required by the problem of interest.

\subsubsection{Transmission conditions}
In addition to~\cref{eq: ns}, \cref{eq: struct_equations} and their associated boundary conditions, the
FSI problem resulting from the embedding of the structural system in $\Omega^F$ is governed by the transmission
conditions
\begin{equation}
\label{eq: velocity_bc}
  v_i = \dot{u}_i \quad \mathrm{on} \; \Sigma,
\end{equation}
and
\begin{equation}
\label{eq: force_bc}
  - p \bm{n} + \bm{\tau} \bm{n} = J^{-1} \bm{F} \bm{S} \bm{F}^T \bm{n} \quad \mathrm{on} \; \Sigma,
\end{equation}
where $\bm{n}$ is the outward unit normal to the deformed configuration of the material interface $\Sigma$,
and $J = \det(\bm{F})$. Since the fluid is assumed to be viscous, additional boundary conditions are
specified on $\Sigma$: these are the adiabatic or isothermal boundary conditions, and the appropriate
boundary conditions for the turbulence model equations when these are presented.

\subsection{Discretization of the governing equations}

\subsubsection{Semi-discretization of the fluid equations}
\begin{figure}
\centering
\includegraphics[scale=0.60]{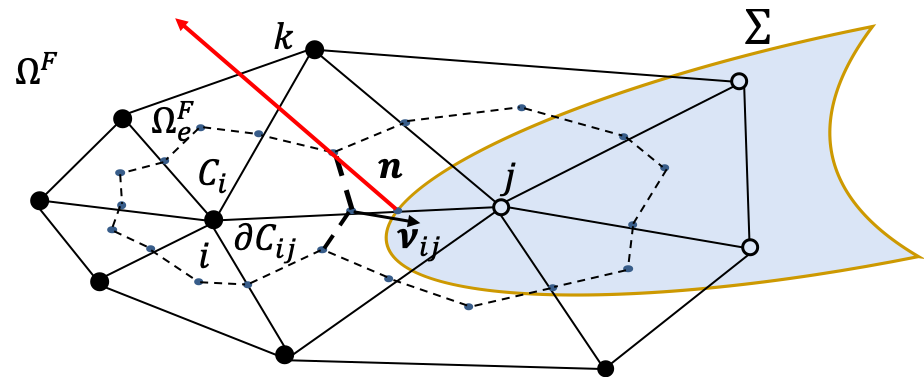}
\caption{Discretization of the fluid domain, dual cell~(control volume) $\mathcal{C}_i$, dual cell boundary
         facet $\partial\mathcal{C}_{ij}$ and its area-weighted outward normal $\bm{\nu}_{ij}$, and fluid-structure
         interface $\Sigma$ and its associated outward normal $\bm{n}$, where the two-material half Riemann
         problem is constructed.}
\label{fig:FIVER}
\end{figure}

Due to the potentially large deformation of the parachute, the Eulerian computational framework with
an immersed (embedded) boundary method \cite{peskin1972flow, mittal2005immersed}) is adopted in the present
study -- specifically, the Finite Volume method with Exact two-phase or two-material Riemann problems~(FIVER)
\cite{farhat2008higher, wang2011algorithms, lakshminarayan2014embedded, huang2018family}. FIVER semi-discretizes
the governing fluid equations~(\cref{eq: ns}) away from the material interface by a hybrid vertex-based Finite
Volume (FV) and Finite Element (FE) method~\cite{farhat1993two}, in which the convective (inviscid) terms
are handled in a vertex-based FV method on an edge-by edge basis, and the viscous terms are handled using
an FE method on an element-by-element basis. 
Let $\Omega_h^F$ denote the discretization of $\Omega^F$ into $N_i$ mesh nodes, and $N_e$ primal elements.
Around each node $i \in \Omega_h^F$, a control volume $\mathcal{C}_i$ is constructed such that
\begin{equation*}
\Omega_h^F = \bigcup \mathcal{C}_i = \bigcup \Omega^F_e,
\end{equation*}
where $\bigcup \mathcal{C}_i$ constitutes a dual mesh, and $\Omega^F_e$ denotes a primal element of this
mesh. The control volume, also called the median-dual control volume, is formed by connecting the centroids,
face, and edge-midpoints of all cells sharing the particular node (see~\cref{fig:FIVER}). Note that
$\partial\mathcal{C}_i$ denotes the boundary of the control volume $\mathcal{C}_i$.

Using the standard characteristic function associated with each control volume $\mathcal{C}_i$, the standard
piecewise linear test function $\psi_i$ associated with each node $i$, and the equivalence between the
two functional spaces generated by the two sets of such functions \cite{farhat1993two}, the weak and
semi-discrete form of \cref{eq: ns} reads
\begin{equation}
\label{eq:nssemi}
  \begin{split}
  \|\mathcal{C}_i\| \frac{d\bm{W}_i}{dt} &+ \sum_{j \in \mathcal{K}(i)} 
  \int_{\partial\mathcal{C}_{ij}}\!\!\! \mathcal{F}\left(\bm{W}\right) \cdot \bm{\nu}_{ij} ds +
  \int_{\partial\mathcal{C}_{i} \cap \partial\Omega^F_\infty} \!\!\!\!\!\mathcal{F}\left(\bm{W}\right) \cdot \bm{n}_\infty ds +\sum_{\Omega^F_e \ni i}\int_{\Omega^F_e} \!\!\!  \nabla{\psi^e_i} \cdot \mathcal{G}\left(\bm{V}, \nabla{\bm{V}}\right) dV  = 0.
\end{split}
\end{equation}
Here, $\|\mathcal{C}_i\|$ denotes the volume of $\mathcal{C}_i$, $\bm{W}_i$ denotes the average values
of $\bm{W}$ in $\mathcal{C}_i$, and $\partial\mathcal{C}_{ij} = \partial\mathcal{C}_i \cap \partial\mathcal{C}_j$
with area-weighted outward normal $\bm{\nu}_{ij}$, $\mathcal{K}(i)$ denotes the set of nodes connected
by an edge to the node $i$, $\partial\Omega^F_\infty$ denotes the far-field boundary of the fluid domain
with outward unit normal $\bm{n}_\infty$, and $\psi^e_i$ denotes the restriction of $\psi_i$ to $\Omega^F_e$. 

The integral of the convective term in \cref{eq:nssemi} away from the fluid-structure interface is approximated
using Roe's (or any other similar) approximate Riemann solver \cite{roe1981approximate} equipped with
a MUSCL technique \cite{van1979towards} and a slope limiter. 
\begin{equation*}
  \int_{\partial\mathcal{C}_{ij}}\!\!\! \mathcal{F}\left(\bm{W}\right) \cdot \bm{\nu}_{ij} ds = \displaystyle \bm{\Phi}_{ij}(\bm{W}_i, 
  \bm{W}_j, \bm{\nu}_{ij}, \hbox{EOS}).
\end{equation*}
The far-field boundary term in \cref{eq:nssemi} is approximated by a standard far-field boundary flux
\cite{steger1981flux,ghidaglia2005normal}. 

The computation of the convective term at the fluid-structure interface is done in two steps, as follows:
\begin{enumerate}
\item for each active node $i$~(in the fluid domain) with an edge $ij$ that intersects the embedded discrete
      surface, a local fluid-structure Riemann problem is formulated and solved at the fluid structure
      interface~\cite{wang2011algorithms} using the structure velocity at the intersection point. This
      furnishes an approximate fluid state $W_i^{\mathcal{R}}$ at the interface. 
\item The convective flux along edge $ij$ in \cref{eq:nssemi} is evaluated by computing the numerical
      flux
      \begin{equation}
      \label{eq:fiver}
        \int_{\partial\mathcal{C}_{ij}}\!\!\! \mathcal{F}\left(\bm{W}\right) \cdot \bm{\nu}_{ij} ds
         = \displaystyle \bm{\Phi}_{ij}(\bm{W}_i, \bm{W}_i^{\mathcal{R}}, \bm{\nu}_{ij}, \hbox{EOS}).
      \end{equation}
\end{enumerate}

The integral of the diffusive flux~(see~\cref{eq:nssemi}) in these elements intersected by the fluid-
structure interface is evaluated using a the ghost fluid method, as follows:
\begin{enumerate}
\item for each active node $i$~(in the fluid domain) within an element $\Omega^F_{e}$ that intersects
      the embedded discrete surface, its neighboring nodes -- which are associated with $\Omega^F_{e}$
      but located on the other side of the fluid-structure interface -- are first identified, then the
      velocity~$\bm{v}$ and temperature~$T$ of these neighboring nodes are populated by linear or constant
      extrapolations.
\item The integral of the diffusive flux is evaluated with a Gaussian quadrature rule, as follows
      \begin{equation}
      \label{eq:nsfem}
        \sum_{\Omega^F_e \ni i} \int_{\Omega^F_e} \nabla{\psi^e_i} \cdot \mathcal{G}\left(\bm{V}, \nabla{\bm{V}}\right) dV \approx
        \sum_{\Omega^F_e \ni i} \|\Omega^F_e\| \nabla{\psi^e_i} \cdot \mathcal{G}\left(\bm{V}^e, \nabla{\bm{V}}^e\right),
      \end{equation}
      where $\|\Omega^F_e\|$ is the volume of the element $\Omega^F_e$ and the terms $\bm{V}^e$ and $\nabla{\bm{V}}^e$
      are defined as
      \begin{equation*}
        \bm{V}^e = \frac{1}{N^e_n}\sum_{k = 1}^{N^e_n} \bm{V}_k \quad \mathrm{and} \quad
        \nabla{\bm{V}}^e = \sum_{k = 1}^{N^e_n} \nabla{\psi^e_k}\bm{V}_k,
      \end{equation*}
      where $N^e_n$ denotes the number of nodes attached to the element $\Omega^F_e$. 
\end{enumerate}

\subsubsection{Semi-discretization of the structural equations}
The governing system of structural equations~(\ref{eq: struct_equations}) is semi-discretized by the FE
method using the total Lagrangian method. The resulting semi-discrete equations of equilibrium can be
written as
\begin{equation}
\label{eq: semi struct}
  \mathbb{M}_h\ddot{\bm{u}}_h + \bm{f}^\mathrm{int}\left(\bm{u}_h, \dot{\bm{u}}_h\right) =  \bm{f}^\mathrm{ext},
\end{equation}
where $\mathbb{M}_h$ denotes the symmetric positive definite FE mass matrix, $\bm{u}_h$ denotes the vector
of semi-discrete structural displacements, $\bm{f}^\mathrm{int}$ and $\bm{f}^\mathrm{ext}$ denote the
vectors of semi-discrete internal and external or flow-induced generalized forces, respectively, and a dot
designates a time derivative.

Specifically, the canopy is modeled using a thin-shell element comprising a membrane triangle with corner
drilling freedoms \cite{alvin1992membrane} and a plate-bending triangle \cite{militello1991first}. Both
membrane stiffness and bending stiffness are considered even though the membrane stiffness is significantly
larger than the bending stiffness. The suspension lines are modeled using Euler-Bernoulli beam elements,
which are fixed at the confluent point. The entry vehicle is modeled as a fixed rigid body, and therefore
does not need to be represented in the structure model.

\subsubsection{Discretization of the governing equations}
Finally, the fluid and structural semi-discretizations~\cref{eq:nssemi,eq: semi struct} are coupled for
FSI simulations using the stability-preserving, second-order, time-accurate, implicit-explicit fluid-structure
staggered solution procedure presented in~\cite{farhat2010robust}. In this coupled time-discretization
algorithm, the semi-discrete fluid subsystem is time-integrated using the second-order three-point implicit
backward difference formula and the semi-discrete structural subsystem is time-integrated using the second-order
explicit central difference time-integration scheme due to the substantial self-contact of the parachute
canopy during its inflation.

\subsection{Homogenized porous wall model}
The porosity of the parachute canopy affects the stability and drag performance of the parachute \cite{kim20062}.
The Ergun equation~\cite{ergun1949fluid} is generally used to simulate fabric porosity \cite{gao2016numerical,
gao2016bnumerical}. However, this equation ignores the compressibility of the flow and hence its validity
is questionable when applied in the supersonic regime. An alternative homogenized porous wall model is proposed
by some of the authors in \cite{huang2019homogenized}, which takes the compressibility effect into consideration and avoids meshing
at the scale of individual pores. This model depends only the void fraction~(or porosity) of the parachute
fabric and the equivalent pore shape, and has been verified using pore-level resolved direct numerical
simulations of typical Mars landing conditions in~\cite{huang2019homogenized}.

Let $\alpha$ denotes the void fraction of the parachute fabric, which is estimated from microtomography
data~\cite{paneraix}, specifically $\alpha = 0.08$ for F-111 nylon. To account for the fabric porosity,
the convective flux through $\partial\mathcal{C}_{ij}$~(see~\cref{eq:fiver}) is modified to be the weighted
average of the fluid-fluid flux and the fluid-structure flux as follows
\begin{equation}
\label{eq:inviscidporous}
  \int_{\partial\mathcal{C}_{ij}} \mathcal{F}(\bm{W}_h) \cdot \bm{\nu}_{ij} ds \approx 
  (1-\alpha)\,\bm{\Phi}_{ij}\big(\bm{W}_{i}, \bm{W}_{i}^{\mathcal{R}}, \bm{\nu}_{ij}\big)
  + \alpha \, \bm{\Phi}_{ij}\big(\bm{W}_{i}, \bm{W}_{j}, \bm{\nu}_{ij}\big).
\end{equation}

The diffusive flux in \cref{eq:nsfem} -- which is evaluated using a ghost fluid method -- is also modified
to account for fabric porosity using a similar technique. Consider the intersected element $\Omega_{e_{ijk}}$
in \cref{fig:FIVER}, let $\bm{V}_j^{g}$ be the populated ghost primitive variables at the node $j$ from
the node $i$ side, and $\bm{V}_j$ be the real primitive variables at the node $j$. Then the quantities
at node $j$ used in \cref{eq:nsfem} are replaced by the weighted average
\begin{equation}
\bm{V}^{ave}_j = \alpha\,\bm{V}_j + (1-\alpha)\,\bm{V}_j^g.
\end{equation}

\subsection{Fluid-structure interaction at the suspension lines subsystem: master-slave kinematics approach}

The presence of suspension lines in the supersonic regime generates shock waves, which have been observed
to disrupt the canopy bow shock and exacerbate parachute oscillation \cite{sengupta2008supersonic}.
However, suspension lines are generally modeled as one-dimensional beam elements due to their large
length-to-diameter ratio, while the fluid subsystem is generally discretized with three-dimensional elements
such as tetrahedra, hexahedra and/or triangular prisms. In this scenario, the enforcement of the FSI
transmission conditions~(\cref{eq: velocity_bc,eq: force_bc}), including the detection of the one-dimensional
elements and the definition of associated geometric characteristics such as surface normals, may be difficult
or even ambiguous in the context of an embedded boundary framework.

\begin{figure}
\centering
\includegraphics[scale=0.50]{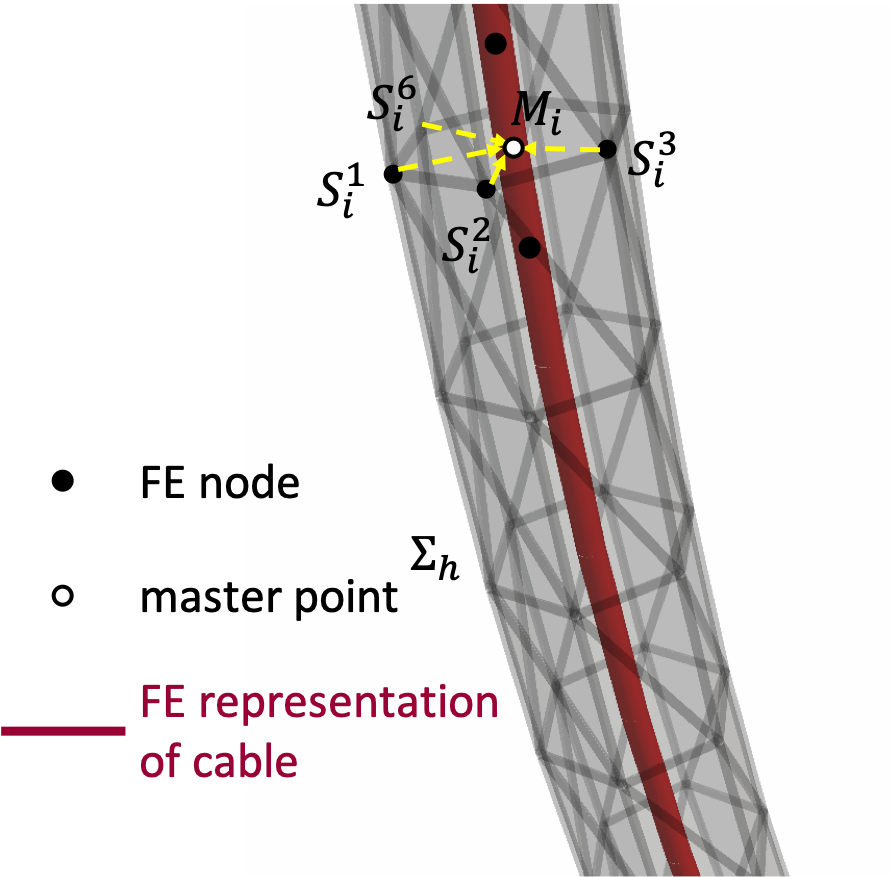}
\caption{Schematics of the master-slave kinematic approach: the discrete surface $\Sigma_h$ representing
         the true geometry of the surface of the cable encloses the topologically 1D FE representation
         of the cable~(red); its nodes are connected to the discretized cable by kinematics constraints
         (yellow) in the master-slave kinematic approach; and $\{S_i^j\}_{j=1}^{n_i}$ denotes the set
         of $n_i$ {\it coplanar} nodes in $\Sigma_h$ whose kinematics are slaved to those of the corresponding
         master {\it point} $M_i$ located at the intersection of the discrete cross section defined by
         the set of nodes $\{S_i^j\}$ and the FE representation of the cable.}
\label{fig:super_beam}
\end{figure}

A master-slave kinematic approach was proposed in~\cite{huang2019embedded}~(see~\cref{fig:super_beam}),
in which the dynamics of the suspension line are captured by the master beam elements, while the real
geometry of the cable is represented within the fluid domain using a slave discrete embedded surface~$\Sigma_h$.
The approach consists of a highly accurate algorithm for computing the embedded surface displacement
based on the beam displacement, and an energy-conserving method for transferring distributed forces and
moments acting on the embedded surface to the beam elements. The load and motion transfer algorithm can be described
as follows:
\begin{enumerate}
\item Pair each slave node $S_i^j$, $j = 1, \cdots, n_i$, with the closest master point $M_i$ on a beam
      element $e_i$ (note that the superscript $j$ highlights the surjective aspect of the function
      $S_i^j \longrightarrow M_i$). The initial distance vector $\bm{d}_i^{j}$ between these two locations
      -- that is, the distance vector at $t = t^0$ -- is defined by
      \begin{equation*}
        \bm{x}^0_{S^j_i} = \bm{x}^0_{M_i} + \bm{d}^j_i.  
      \end{equation*}
\item At each time step, compute the displacement $\bm{u}_{S^j_i}$ and velocity $\dot{\bm{u}}_{S^j_i}$
      of the slave node $S_i^j$ as follows
      \begin{equation*}
        \bm{u}_{S^j_i} = \bm{u}_{M_i} + \mathcal{R}(\bm{\theta}_{M_i}) \bm{d}^j_i - \bm{d}^j_i  ~~~~~~\mathrm{and}~~~~~~~~
        \dot{\bm{u}}_{S^j_i} = \dot{\bm{u}}_{M_i} + \bm{\omega}_{M_i} \times \mathcal{R}(\bm{\theta}_{M_i})\bm{d}^j_i, 
      \end{equation*}
      where $\bm{u}_{M_i}$ and $\bm{\theta}_{M_i}$ denote the interpolated displacement and rotation
      vectors at the point $M_i$, $\dot{\bm{u}}_{M_i}$ and ${\bm{\omega}}_{M_i}$ are the interpolated
      velocity and angular velocity vectors at the point $M_i$, and $\mathcal{R}$ is the rotation matrix
      at $M_i$ which depends on $\bm{\theta}_{M_i}$. 
\item Compute the force vector $\bm{f}_{S^j_i}$ at each slave node $S_i^j$ as follows
      \begin{equation*}
        \bm{f}_{S^j_i} = \int_{\Sigma_h} \Big( -p \bm{n} + \bm{\tau} \bm{n}\Big)\phi_{S^j_i}\,d\Sigma_h,
      \end{equation*}
      where $p$ and $\bm{\tau}$ denote the pressure and viscous stress tensor of the flow at this time
      step, $\bm{n}$ denotes the outward normal to $\Sigma_h$ at this time step, and $\phi_{S^j_i}$ denotes
      a {\it local} shape function associated with the node $S^j_i \in \Sigma_h$. The force vector $\bm{f}_{M_i}$
      and moment vector $\bm{m}_{M_i}$ at the point $M_i$ are computed as follows
      \begin{equation*} 
        \bm{f}_{M_i} = \sum_{j = 1}^{n_i}\bm{f}_{S^j_i}  ~~~~~~~\mathrm{and}~~~~~~~~
        \bm{m}_{M_i} = \sum_{j = 1}^{n_i}\mathcal{R}(\bm{\theta}_{M_i})\,\bm{d}^j_i \times \bm{f}_{S^j_i}.
      \end{equation*}
      Finally, the generalized force and moment vectors acting on FE nodes of the beam element are assembled
      by the load transfer method presented in~\cite{farhat1998load}.
\end{enumerate}
Hence, both the flow-induced forces on the cable and the effect of the structural dynamic response of
the cable on the nearby flow are taken into account.

\subsection{Local contact algorithm}

Massive self-contact occurs during parachute inflation, and must be accounted for in order to accurately
simulate the parachute inflation dynamics. Moreover, self-penetration causes numerical instability for
the fluid solver in the Eulerian computational framework. In this work, the self-contact law for the canopy is enforced
by Lagrangian multiplier method using the Algorithms for Contact in a Multiphysics Environment~(ACME)
library~\cite{heinstein2004acme}. Furthermore, to accelerate the contact detection and relieve the algorithmic difficulties
associated with the reliable detection of interactions for thin two-sided contact surfaces, the contact is
enforced between each pair of neighboring gores and only from the outside~(i.e., one sided contact).
The validity of this simplifying assumption was verified \emph{a posteriori} by carefully inspecting
the solution for self-penetration.

\section{Simulation results}
\label{sec:validation}

The computational framework summarized above is implemented in the massively parallel AERO Suite 
\cite{farhat2003application}. It has been verified and validated for several large-scale, highly nonlinear
applications associated with marine engineering, aerospace engineering~\cite{farhat2013dynamic}, and flapping wings
\cite{lakshminarayan2014embedded}. It is applied here to simulate the DGB parachute inflation described
in \cref{sec:setup}. Overall, four scenarios are considered in order to investigate different factors
that potentially affect the parachute inflation dynamics in the Martian atmosphere. These are listed below
\begin{enumerate}
\item Scenario 1: both the interaction of the suspension lines with the flow, and the contribution of
      bulk viscosity of the fluid medium are neglected. The inflow conditions are the supersonic free-stream
      Mars atmospheric conditions corresponding to the mortar fire event of the Curiosity mission given
      by
      \begin{equation*}
        \rho^{F}_{\infty} = 0.0067 \; \mathrm{kg}/\mathrm{m}^3,
        \quad p_{\infty} = 260 \; \mathrm{Pa}, \quad \mathrm{and}
        \quad \mathsf{Ma}_{\infty} = 1.8.
      \end{equation*}
      The Reynolds number based on the canopy diameter and the inflow conditions is $4.06 \times 10^6$.
\item Scenario 2: the interaction of the suspension lines with the flow are accounted for, but the contribution
      of the bulk viscosity of the fluid medium is neglected. The inflow conditions are the same as for
      Scenario 1.
\item Scenario 3: both the interaction of the suspension lines with the flow, and the contribution of
      the bulk viscosity of the fluid medium, are accounted for. The inflow conditions are the same as for
      Scenario 1.
\item Scenario 4: both the interaction of the suspension lines with the flow, and the contribution of 
      the bulk viscosity of the fluid medium are accounted for, as for Scenario 3. Slightly different
      inflow conditions corresponding to the line stretch event of the Curiosity mission are used however -- specifically,
      \begin{equation*}
        \rho^{F}_{\infty} = 0.0060 \; \mathrm{kg}/\mathrm{m}^3,
        \quad p_{\infty} = 244.4 \; \mathrm{Pa}, \quad \mathrm{and}
        \quad \mathsf{Ma}_{\infty} = 1.74.
      \end{equation*}
\end{enumerate}

The background computational fluid domain is initially discretized by a mesh composed of Kuhn simplices
\cite{stevenson2008completion, borker2019mesh}. This initial tetrahedral mesh contains $2,778,867$ vertices
and $16,308,672$ tetrahedra. Adaptive mesh refinement (AMR)~\cite{borker2019mesh} based on newest vertex
bisection~\cite{mitchell1988unified, stevenson2008completion} is used, enabling the boundary layer and
flow features to be efficiently tracked using a wall distance estimator and a Hessian error indicator,
respectively. However, fully resolving the boundary layer of the cable subsystem is generally unaffordable,
especially when the cable has large length-to-diameter ratio. To obtain a minimally acceptable resolution,
the doubly-intersected edge criterion~\cite{huang2019embedded} is applied on each suspension line. Specifically,
when an edge of the fluid mesh is intersected twice by the cable's outer surface -- which indicates that
the cable is under-resolved in this proximity -- the edge is selected for refinement and subsequently
bisected. The characteristic mesh sizes near the entry vehicle, suspension lines and the canopy are
$2.5$~cm, $3$~mm, and $5$~cm, respectively, while the mesh size in the wake and near the shock is $10$~cm.
During the mesh adaptation, the maximum numbers of vertices and tetrahedra reach about $42,000,000$ and
$238,000,000$, respectively.

The canopy of the DGB parachute consists of band gores and disk gores. These are discretized here by $279,025$
geometrically nonlinear ANDES thin shell elements~\cite{militello1991first}. Each one of the 80 suspension lines
is discretized by $500$ geometrically nonlinear beam elements. The cross-section geometry
of each suspension line is assumed to be circular and is represented as a hexagon.

The fluid-structure coupling time-step is initialized to $\Delta t_{F/S} = 10^{-5}$~s, but is able to vary
during the simulation to preserve stability of the conditionally-stable explicit structural time-integration
scheme.

\begin{figure}[!htbp]
  \centering
  \begin{subfigure}[b]{0.33\textwidth}
    \centering
    \includegraphics[scale=0.095]{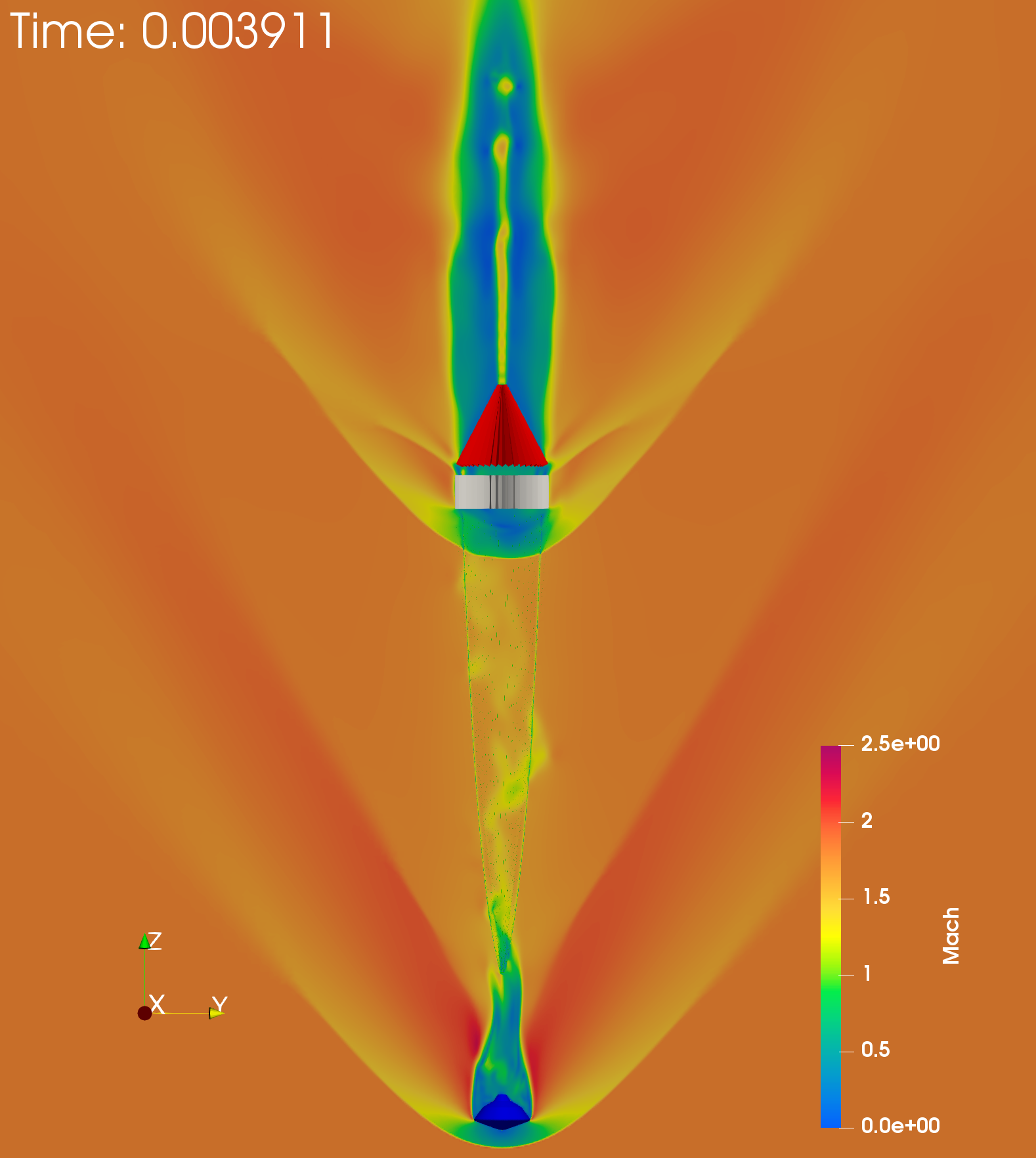}
    \caption{}
  \end{subfigure}
  \begin{subfigure}[b]{0.33\textwidth}
    \centering
    \includegraphics[scale=0.095]{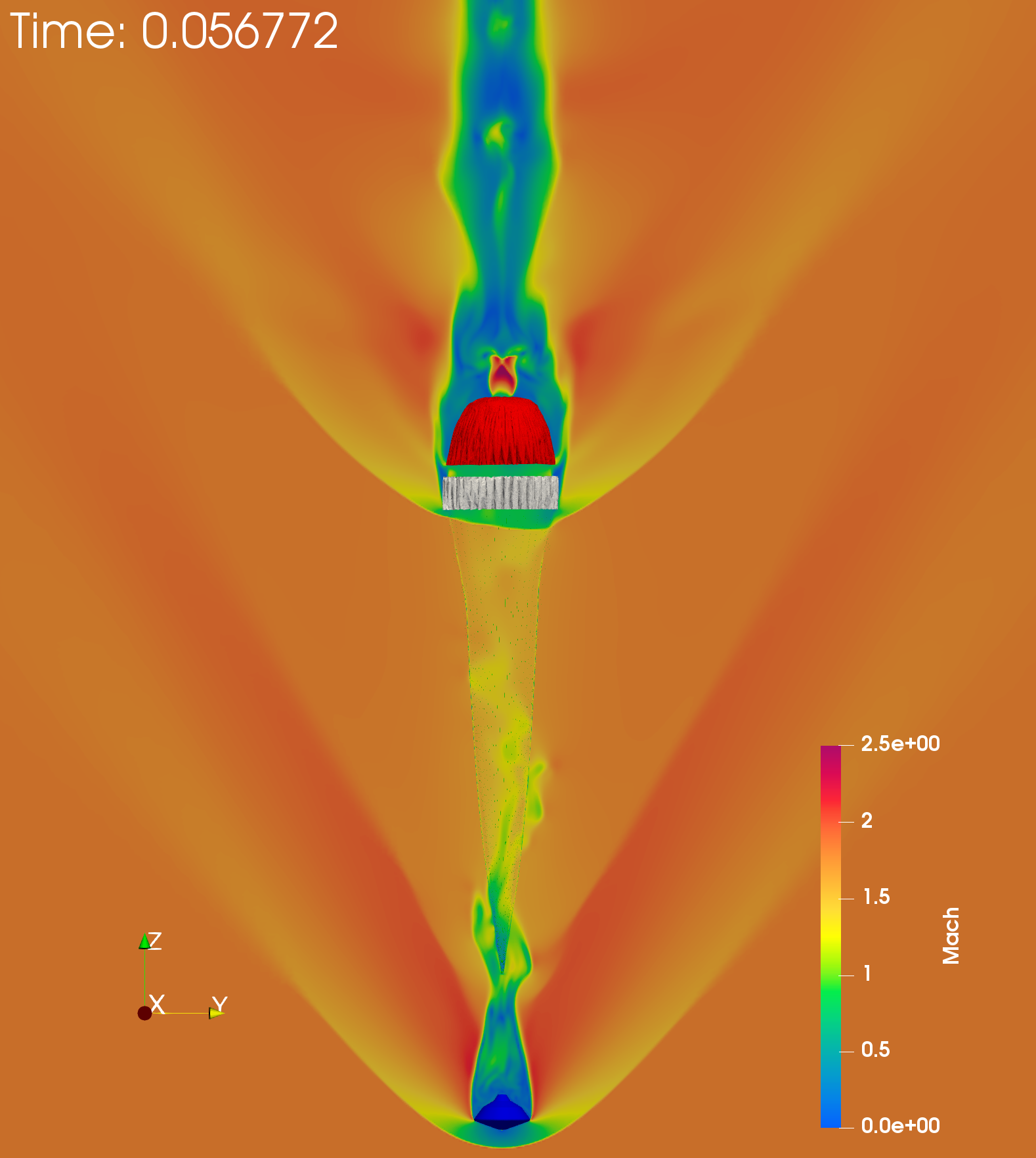}
    \caption{}
  \end{subfigure}
  \begin{subfigure}[b]{0.33\textwidth}
    \centering
    \includegraphics[scale=0.095]{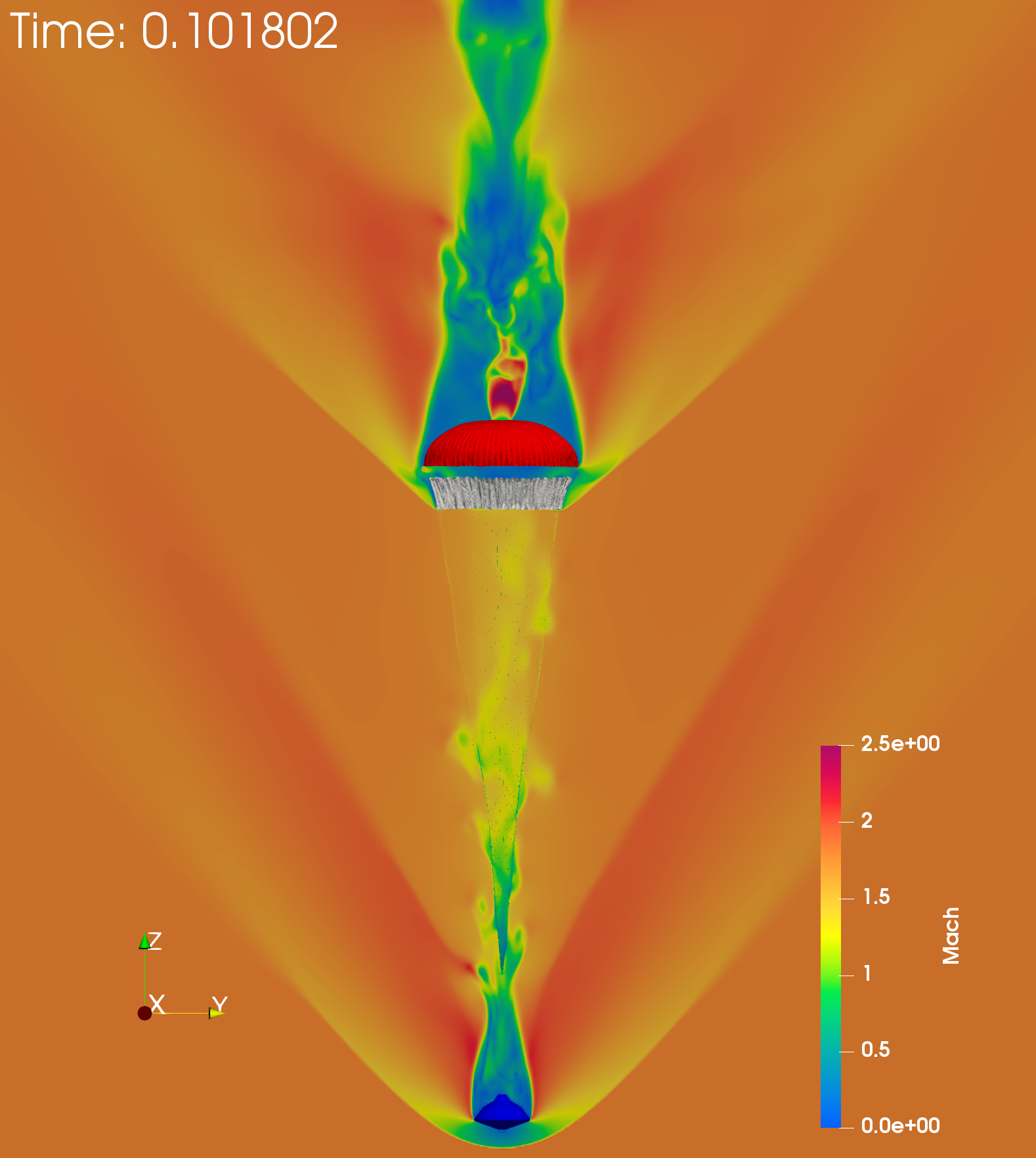}
    \caption{}
  \end{subfigure}    
    \begin{subfigure}[b]{0.33\textwidth}
    \centering
    \includegraphics[scale=0.095]{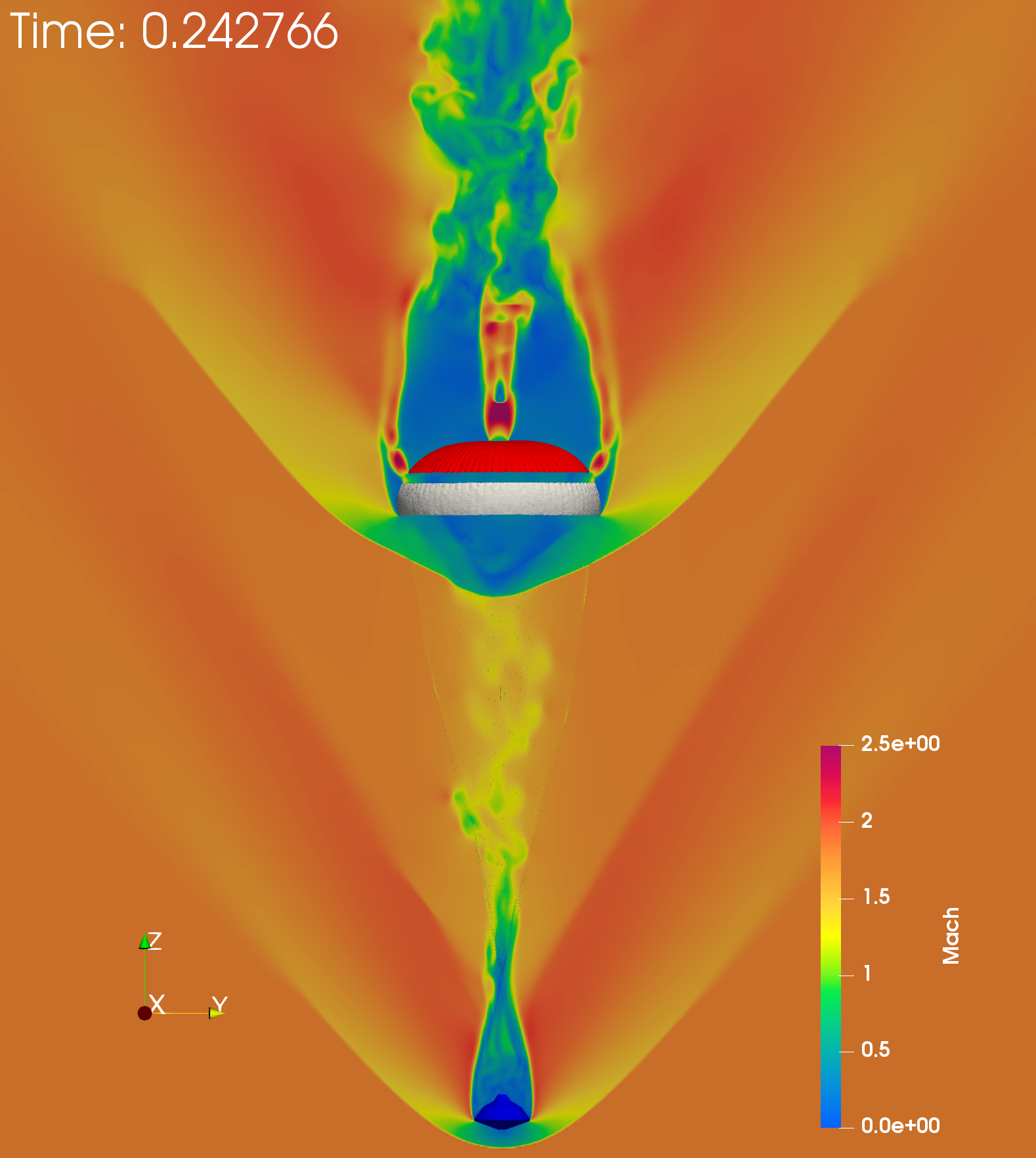}
    \caption{}
  \end{subfigure}
  \begin{subfigure}[b]{0.33\textwidth}
    \centering
    \includegraphics[scale=0.095]{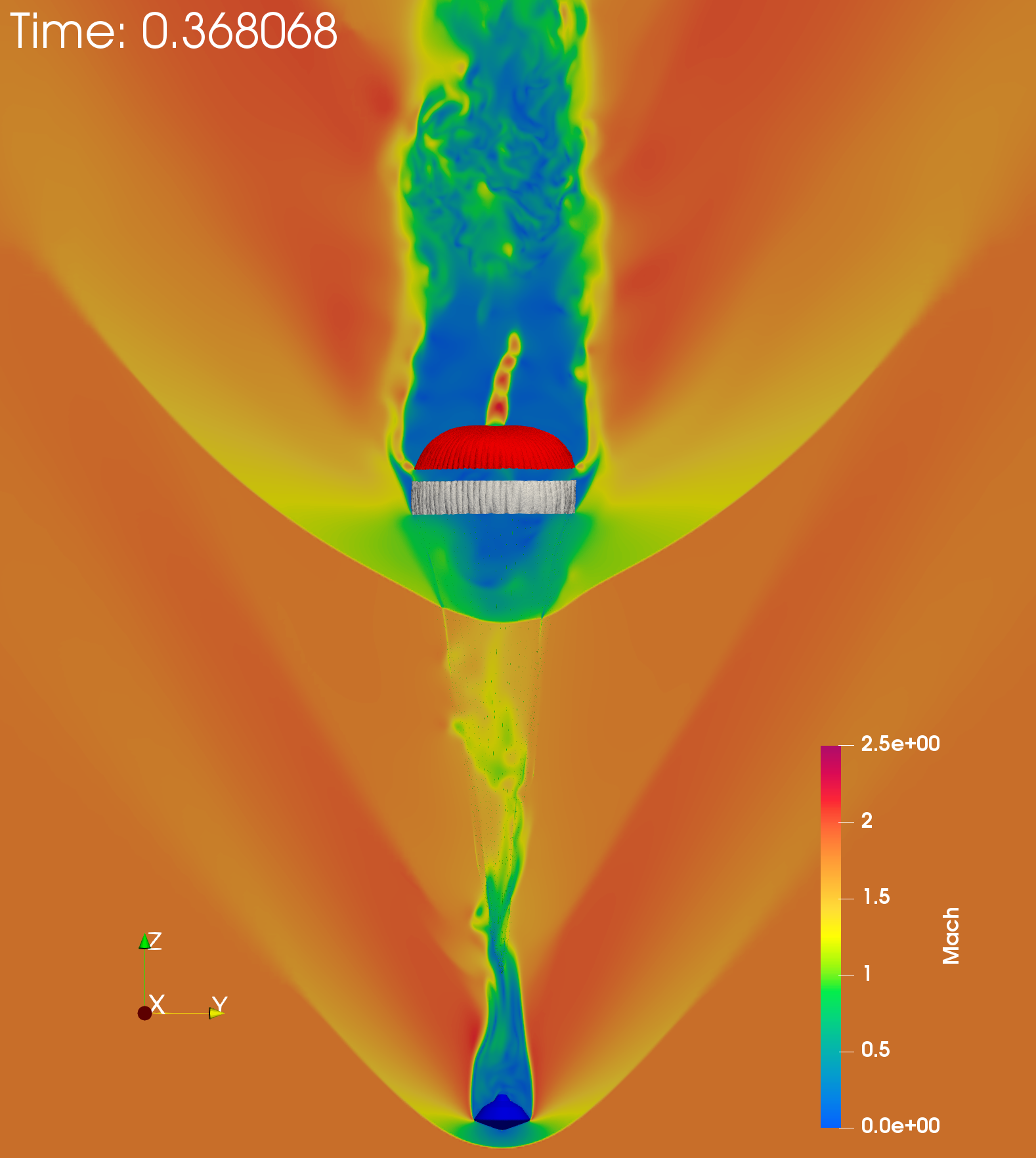}
    \caption{}
  \end{subfigure}
  \begin{subfigure}[b]{0.33\textwidth}
    \centering
    \includegraphics[scale=0.095]{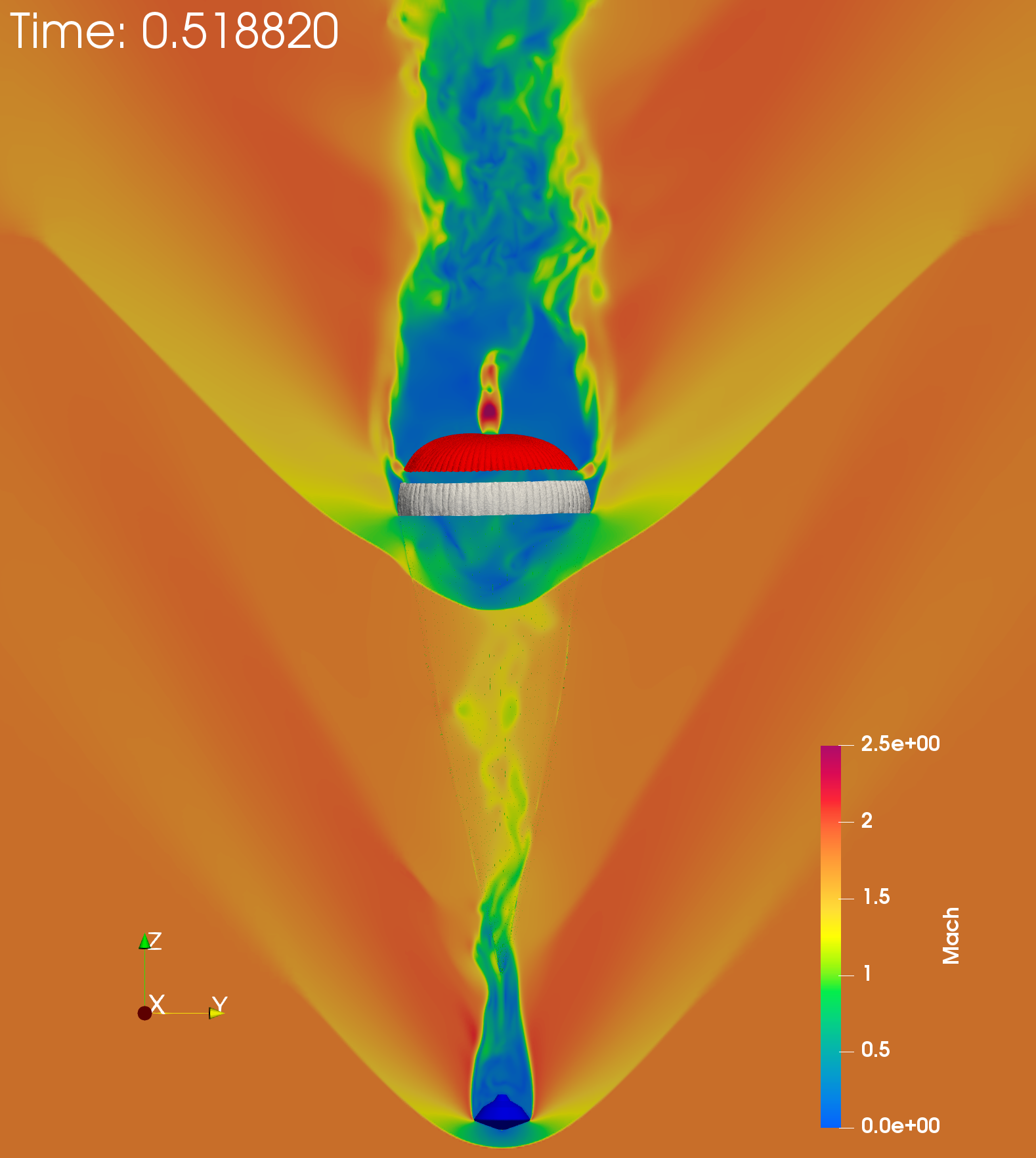}
    \caption{}
  \end{subfigure} 
  \caption{AMR-enabled, LES-based, simulation of the supersonic parachute inflation dynamics problem:
           snapshots of the time-evolution of the parachute and the flow Mach number field~(Scenario 2).}
\label{fig:dgb-c2}
\end{figure}

\begin{figure}[!htbp]
  \centering
  \begin{subfigure}[b]{0.54\textwidth}
    \centering
    \includegraphics[scale=0.65]{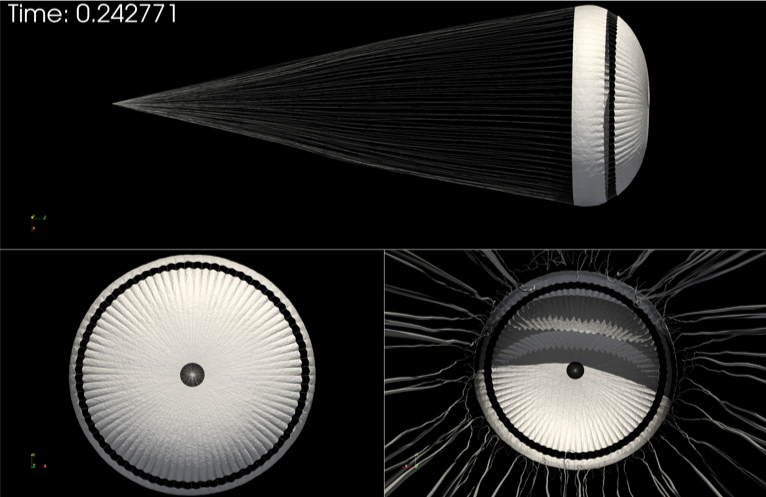}
    \caption{}
  \end{subfigure}
  \begin{subfigure}[b]{0.44\textwidth}
    \centering
    \includegraphics[scale=0.65]{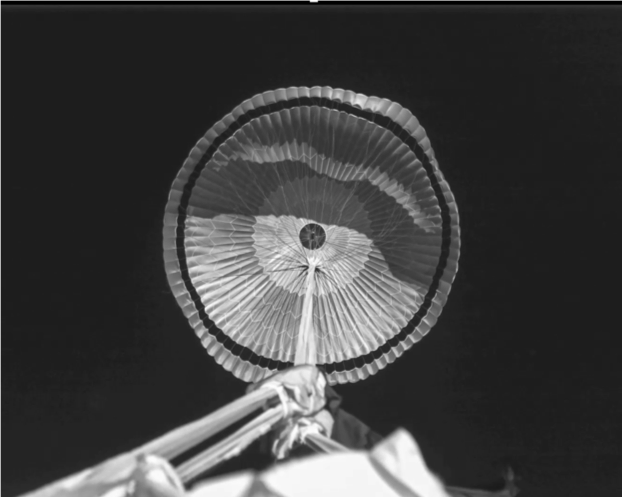}
    \caption{}
  \end{subfigure}
  \caption{Simulated parachute structural state~(Scenario 2) at its full inflation moment~(a), and counterpart
           test-captured (deep space flight test) parachute structural state using a high-speed camera~(b).}
    \label{fig:dgb-c2-stru}
\end{figure}

\Cref{fig:dgb-c2} graphically depicts the time-evolution of the parachute deployment and the flow Mach
number field for Scenario 2. The inflow is from the bottom of the computational fluid domain. The supersonic
inflow first forms a steady bow shock in front of the entry vehicle, then slows down into the subsonic regime
behind it, and finally recovers the supersonic regime and forms an unsteady bow shock in front of the canopy.
At the beginning of the FSI, the bow shock in the front of the canopy moves toward the canopy~(see \cref{fig:dgb-c2}-a
and \cref{fig:dgb-c2}-b), the disk part of the canopy begins inflation and reaches full inflation at
about $t = 0.1$~s~(see \cref{fig:dgb-c2}-c). A supersonic jet (choked flow) appears when the high pressure
flow enveloped by the canopy ejects through the disk vent, which intensively interacts with the turbulent
wakes behind the canopy. The high pressure flow inside the canopy inflates the band part at around $t = 0.24$~s
(see \cref{fig:dgb-c2}-d), which leads to the peak load on the parachute. Following that, the canopy starts
to ``breathe'' and forms several partial collapse cycles. \Cref{fig:dgb-c2-stru}-a visualizes the state
of the FE structural model including the suspension lines at the moment of the full inflation, and also
includes a recent deep space flight test photo of this state for the purpose of qualitative validation
(see \Cref{fig:dgb-c2-stru}-b).

\subsection{Drag performance analysis}

\begin{figure}
\centering
\includegraphics[scale=0.30]{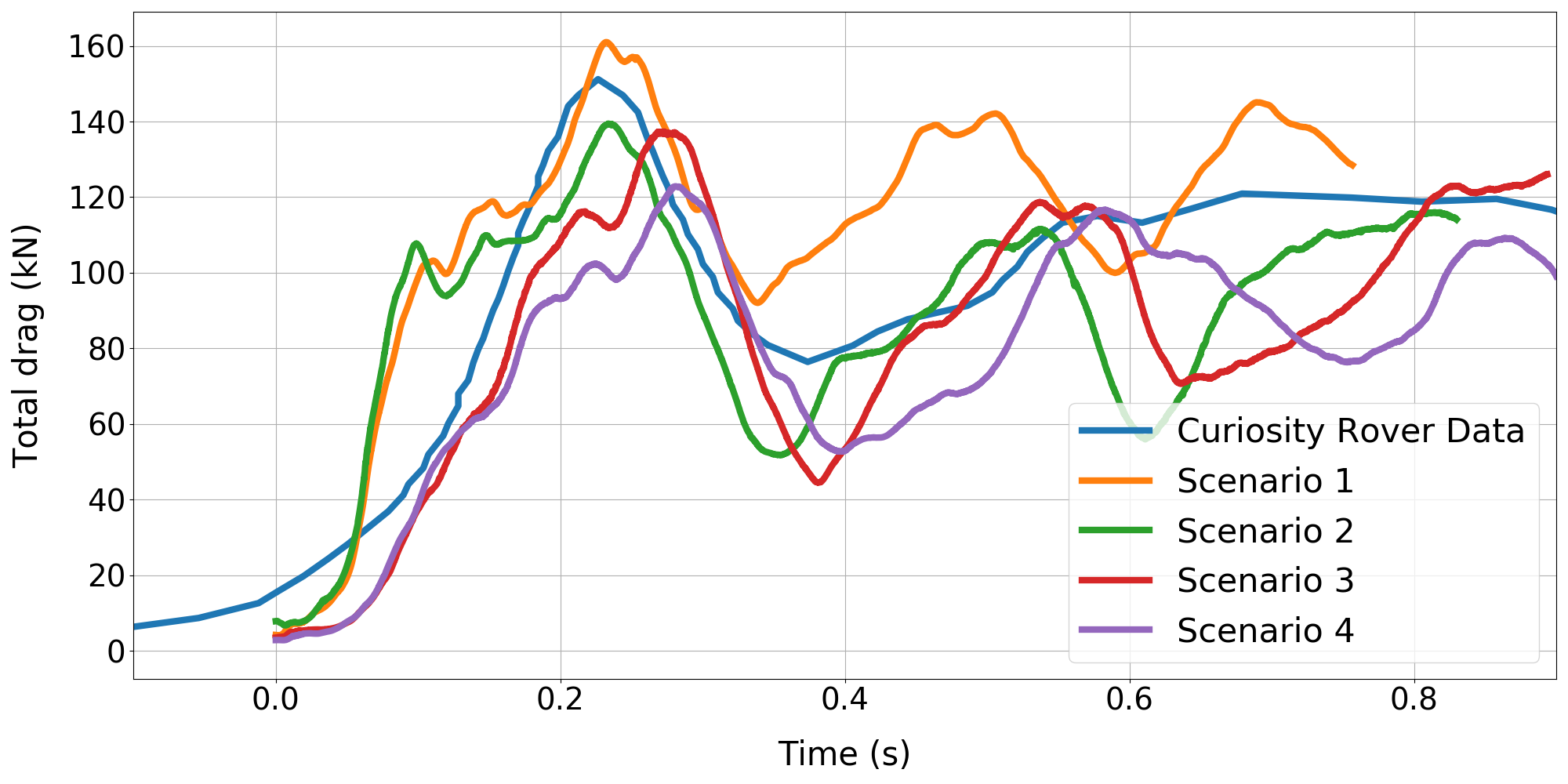}
\caption{Measured and predicted time-histories of the total drag force generated by both the parachute
         canopy and its suspension lines during the parachute inflation.}
\label{fig:Drag_all}
\end{figure}

\Cref{fig:Drag_all} reports the time-histories of the measured and predicted drag forces during the Curiosity
Mars landing. Note that each force time-history includes contributions of both the suspension lines and
the canopy. Although the flight-test data is collected during the first $4$~s following the mortar fire,
\cref{fig:Drag_all} focuses on the time-interval following the suspension line stretch which includes the
peak drag. The reported time-histories show that the parachute inflates rapidly, generates a total drag
force of roughly $153.8$~kN, then undergoes several partial collapse cycles. 

\begin{figure}[!htbp]
  \centering
  \begin{subfigure}[b]{0.49\textwidth}
    \centering
    \includegraphics[scale=0.29]{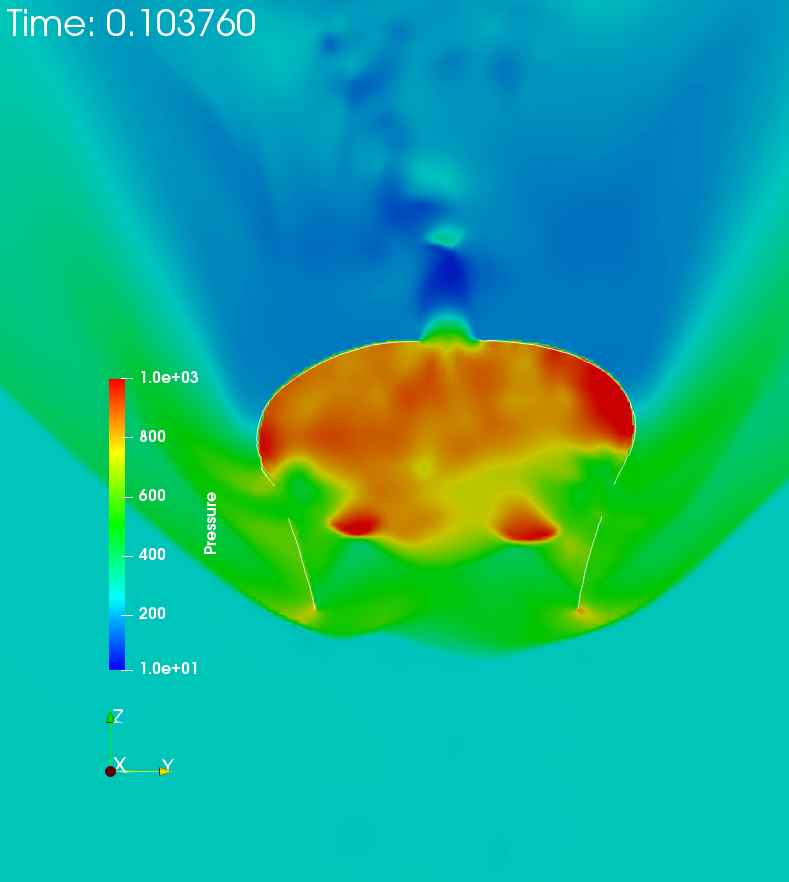}
    \caption{}
  \end{subfigure}
  \begin{subfigure}[b]{0.49\textwidth}
    \centering
    \includegraphics[scale=0.29]{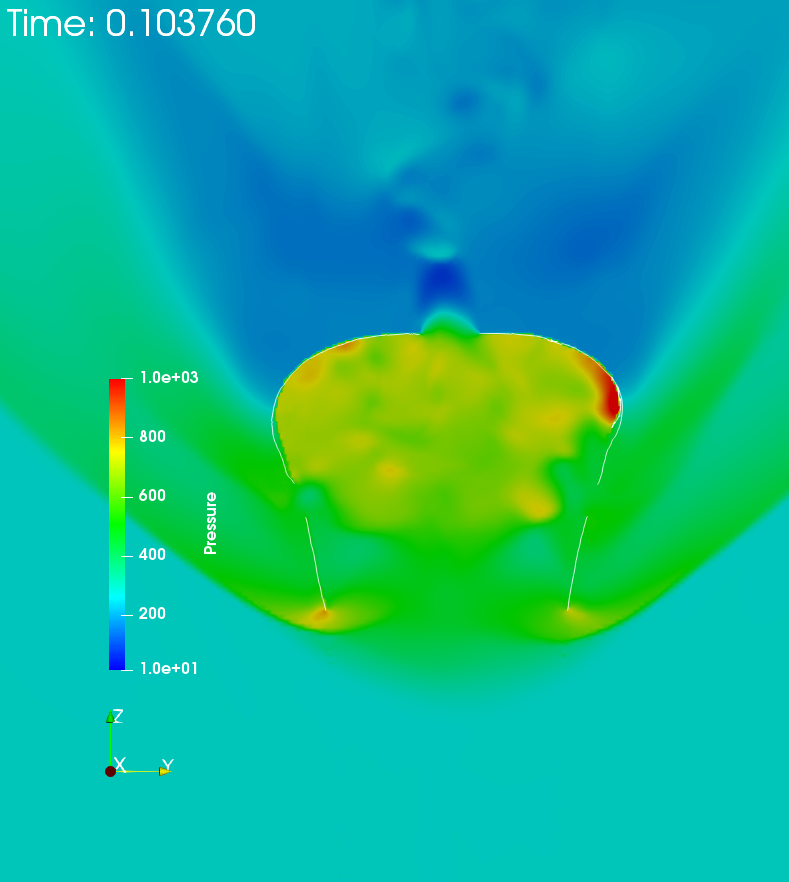}
    \caption{}
  \end{subfigure}\\
  \begin{subfigure}[b]{0.49\textwidth}
    \centering
    \includegraphics[scale=0.29]{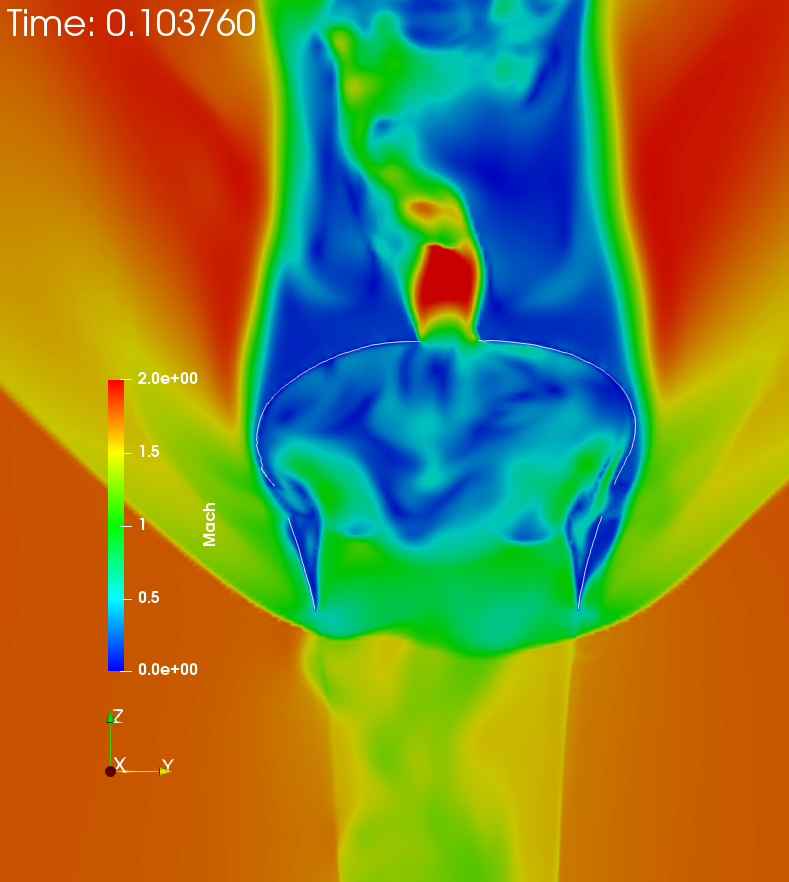}
    \caption{}
  \end{subfigure}    
    \begin{subfigure}[b]{0.49\textwidth}
    \centering
    \includegraphics[scale=0.29]{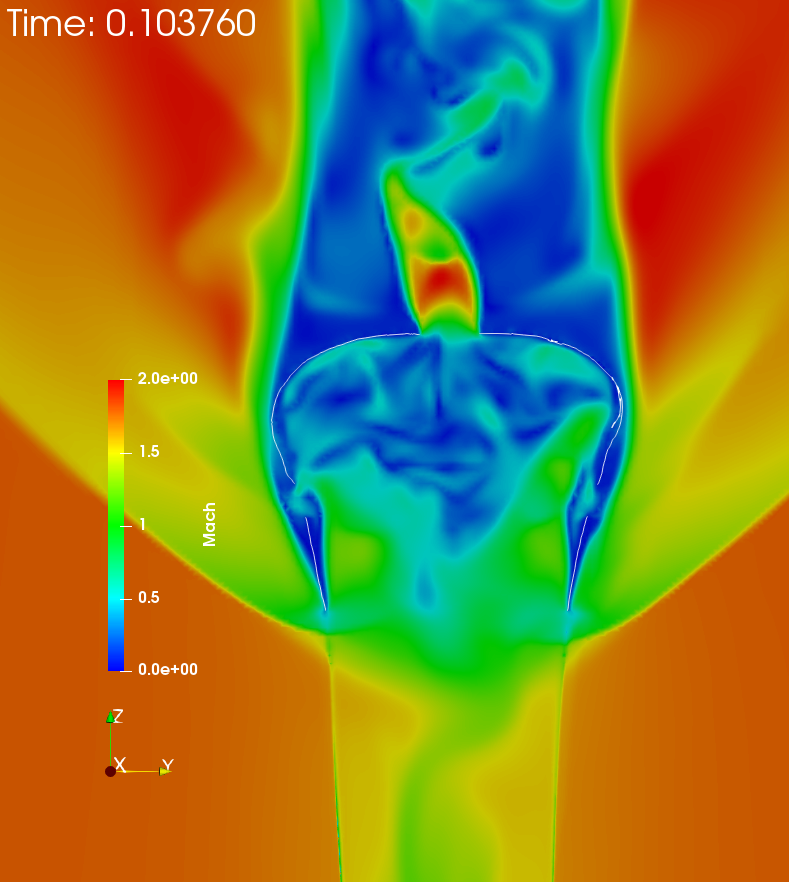}
    \caption{}
  \end{subfigure}
  \caption{Pressure profiles~(top) and Mach profiles~(bottom) from Scenario 2~(left) and Scenario 3~(right)
           of the parachute inflation at $0.1$~s, which corresponds to the inflation of the disk part.}
    \label{fig:dgb-bulk}
\end{figure}

The visualization of the simulated Mach field results~(see \cref{fig:dgb-c2}-b and \cref{fig:dgb-c2}-e)
shows that the suspension lines create disturbances ahead of the canopy that disrupt the parachute bow shock,
promote the mixing of the high pressure and low-pressure flows, and reduce the pressure inside the canopy.
Consequently, and as can be inferred from comparison between drag force histories of Scenario 1~(without
fluid suspension line interaction) and Scenario 2~(with fluid suspension line interaction) in \cref{fig:Drag_all},
the interactions of the suspension lines with the nearby high-speed flows reduce the peak drag and the
overall drag performance. 
\Cref{fig:Drag_all} also shows that as far as the peak of the total drag force is concerned, the effect
of the bulk viscosity of the flow medium is rather weak. However, it is significant near the bow shock
and jet streams through the vent and the gap, where the volumetric dilatation rate $\nabla \cdot \bm v$
is large. The bulk viscosity effect mainly appears in the drag force histories at about $0.1$~s, when
the disk part is inflated along with the approaching of the bow shock. The pressure profiles and Mach
profiles of Scenario 2~(without bulk viscosity) and Scenario 3~(with bulk viscosity) at about $0.1$~s
are depicted in \cref{fig:dgb-bulk}. For Scenario 3, the large bulk viscosity smears the moving bow shock
(see \cref{fig:dgb-bulk}-d). Hence, the pressure inside the parachute canopy is lower than for Scenario 2. 
The comparison between Scenario 3 and Scenario 4 reveals the sensitivity of the present parachute inflation
simulations with respect to the (uncertain) initial flow conditions.

Overall, the peak drag predictions are in good agreement with the flight test data, even though in all
performed AERO Suite simulations, the forebody is assumed to be rigid and stationary, the trim angle
is arbitrarily set to zero, and the deceleration process is not accounted for. To the best of our knowledge,
the parachute deployment simulations reported here are the first FSI simulations to match the Mars landing
data of Curiosity. 

\subsection{Material failure analysis}
Understanding/predicting the onset of the canopy failure that has been observed in several flight tests is
another major objective of this research effort. In all numerical simulations discussed above, the canopy
material is modeled as a St. Venant-Kirchhoff elastic material, and the von Mises stress is used as the
material failure indicator for a preliminary failure analysis, as will be shown below.

\begin{figure}
\centering
  \centering
  \begin{subfigure}[b]{0.49\textwidth}
    \centering
    \includegraphics[scale=0.47]{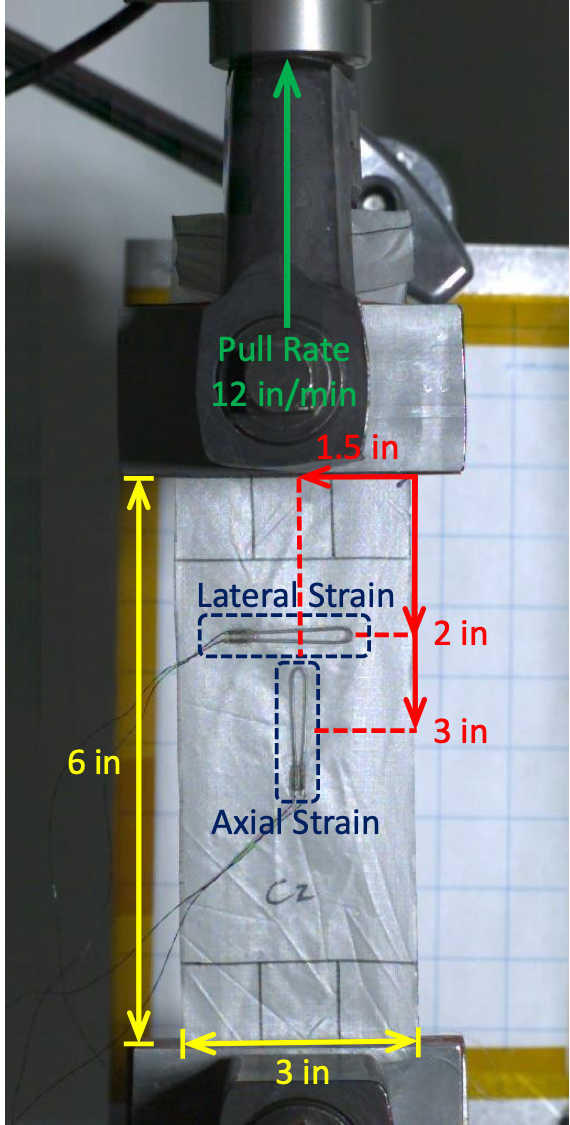}
    \caption{}
  \end{subfigure}
  \begin{subfigure}[b]{0.49\textwidth}
    \centering
    \includegraphics[scale=0.2]{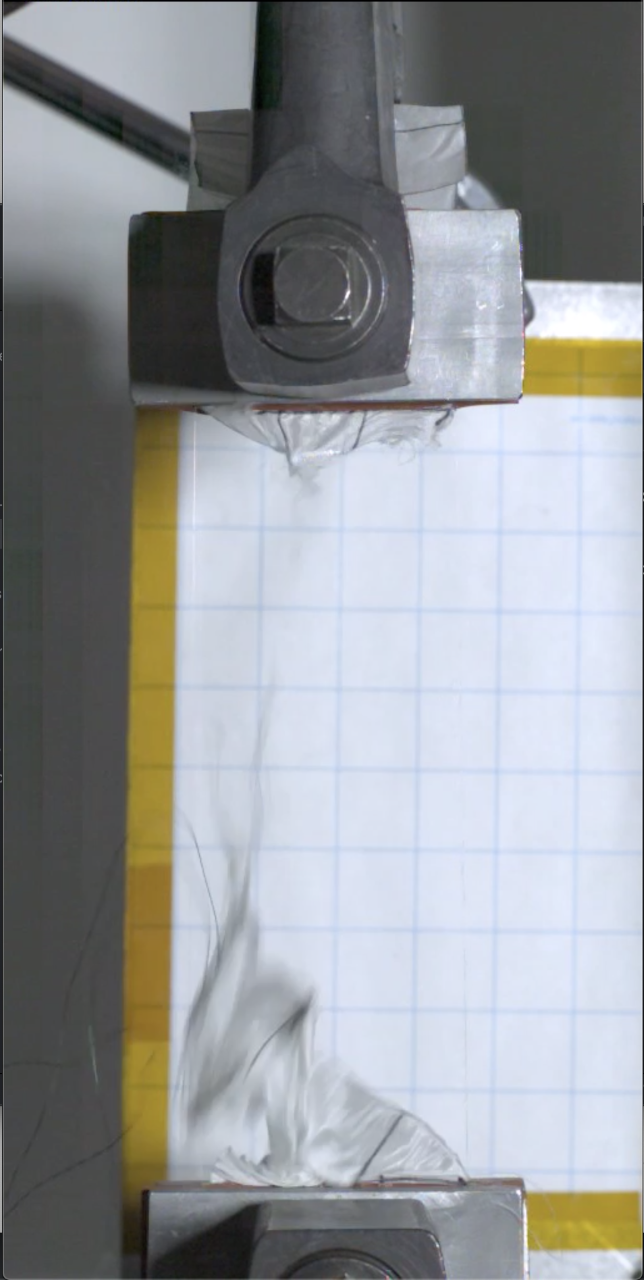}
    \caption{}
  \end{subfigure}
\caption{Uniaxial tensile test of the nylon material: experimental setup (a) and failure moment (b).}
\label{fig:coupon}
\end{figure}

To find the von Mises yield stress at failure, several uniaxial tensile tests of nylon coupons were
conducted\footnote{The tests were performed on a representative fabric coupon which is close to but
not necessarily the same as that used for the Curiosity rover. These tests were also performed to assess
a new type of strain gauge, which may create a local stiffness and alter the physical characteristics
of the fabric at the point of contact, due to the adhesive.}.
The specification of the corresponding experimental setup is illustrated in \cref{fig:coupon}-a.
The coupon of size $3$~in by $6$~in is clamped on its bottom edge and pulled from its top edge with a pull
rate of $12$~in/min until failure. An axial strain sensor is located at the center of the coupon. The
coupon is torn to two pieces at about $t = 7.5$~s, with the point of failure located close to the center
of the coupon (see \cref{fig:coupon}-b). 

\begin{figure}
\centering
\includegraphics[scale=0.30]{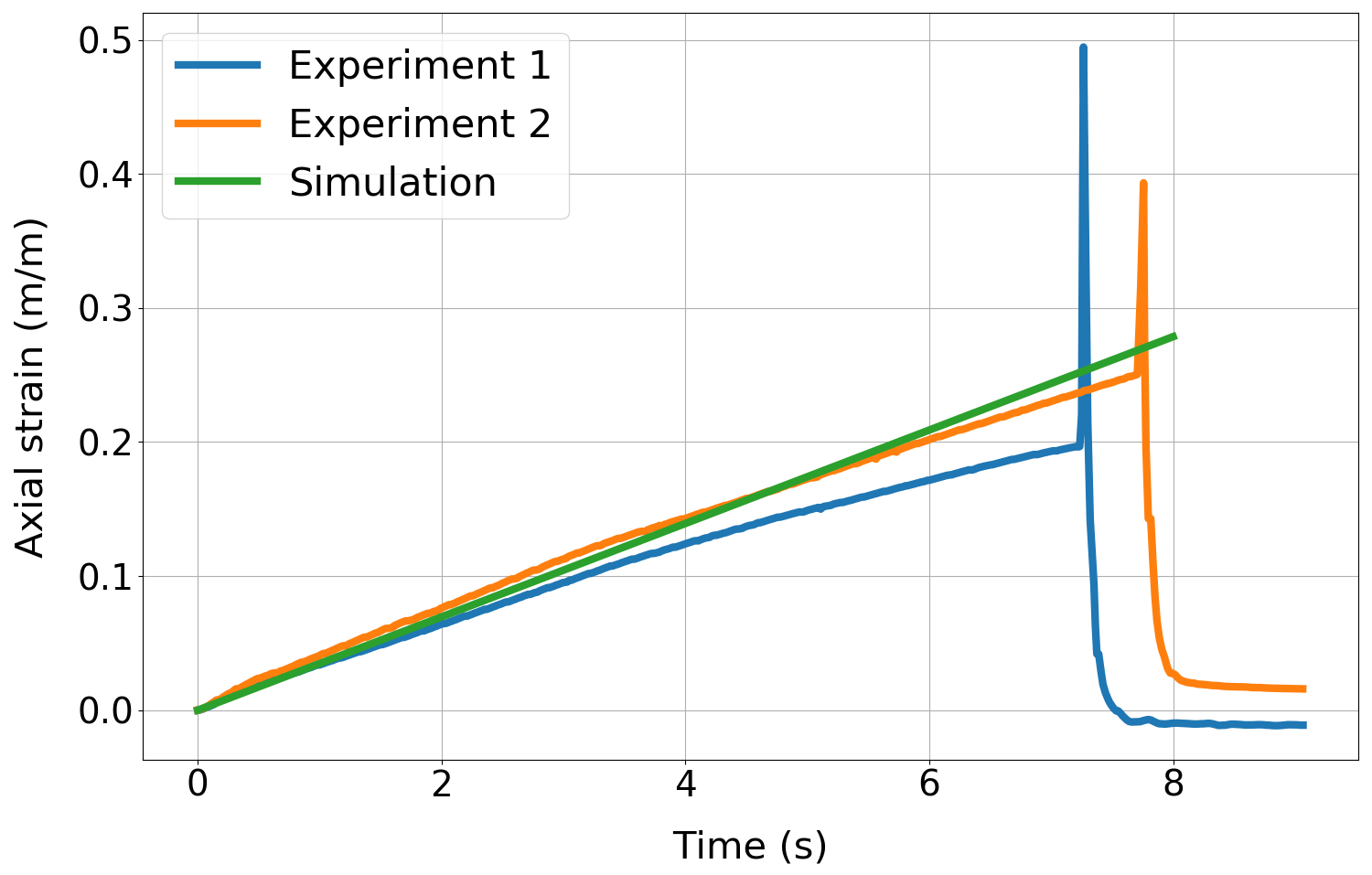}
\caption{Simulation results of the nylon material uniaxial tensile test: axial strain at the center.}
\label{fig:coupon_strain}
\end{figure}

\begin{figure}
\centering
\includegraphics[scale=0.25]{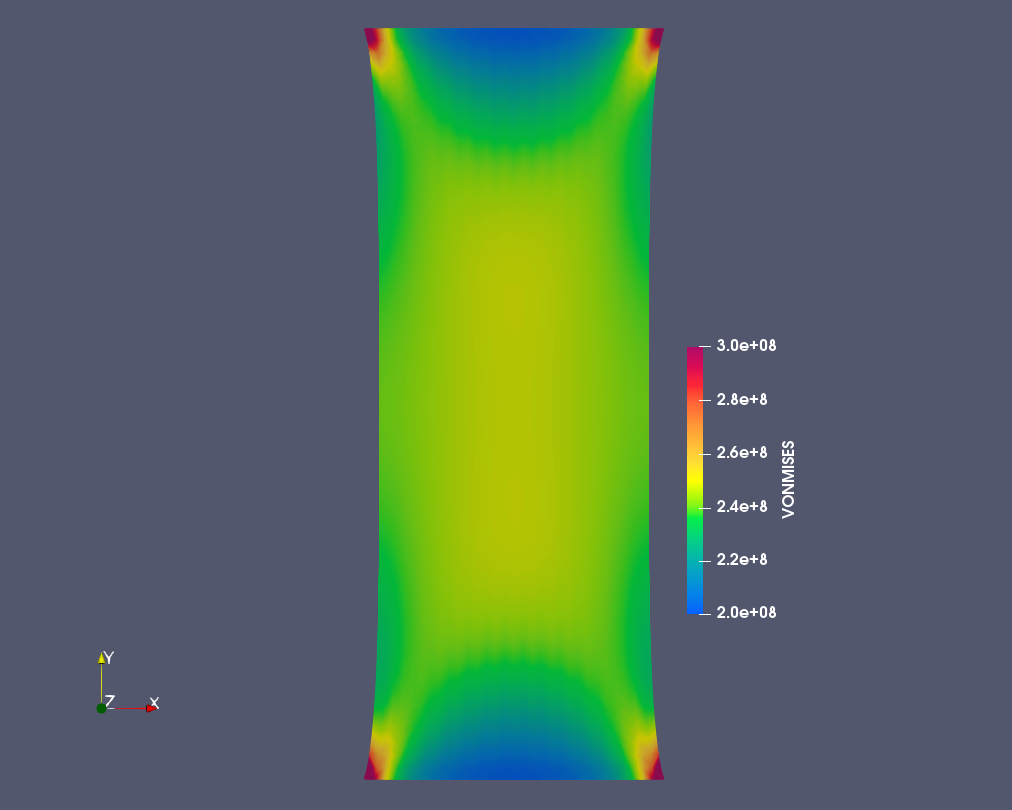}
\caption{Simulation results of the nylon material uniaxial tensile test: von Mises stress distribution at the failure moment.}
\label{fig:coupon_vm}
\end{figure}

A numerical simulation of this uniaxial test is conducted using the same specifications as above. The
measured and predicted axial strains are reported in~\cref{fig:coupon_strain}: the predicted axial strain
can be seen to match well with the experimental data before material failure. The computed von Mises
stress at the point of failure around $t = 7.5$~s was found to be about $2.5 \times 10^8$~Pa (see \cref{fig:coupon_vm}).
Note the stress concentrations at the four corners are artificial as they are due to the inexact modeling
of the boundary conditions, and were therefore neglected.

\begin{figure}
\centering
\includegraphics[scale=0.30]{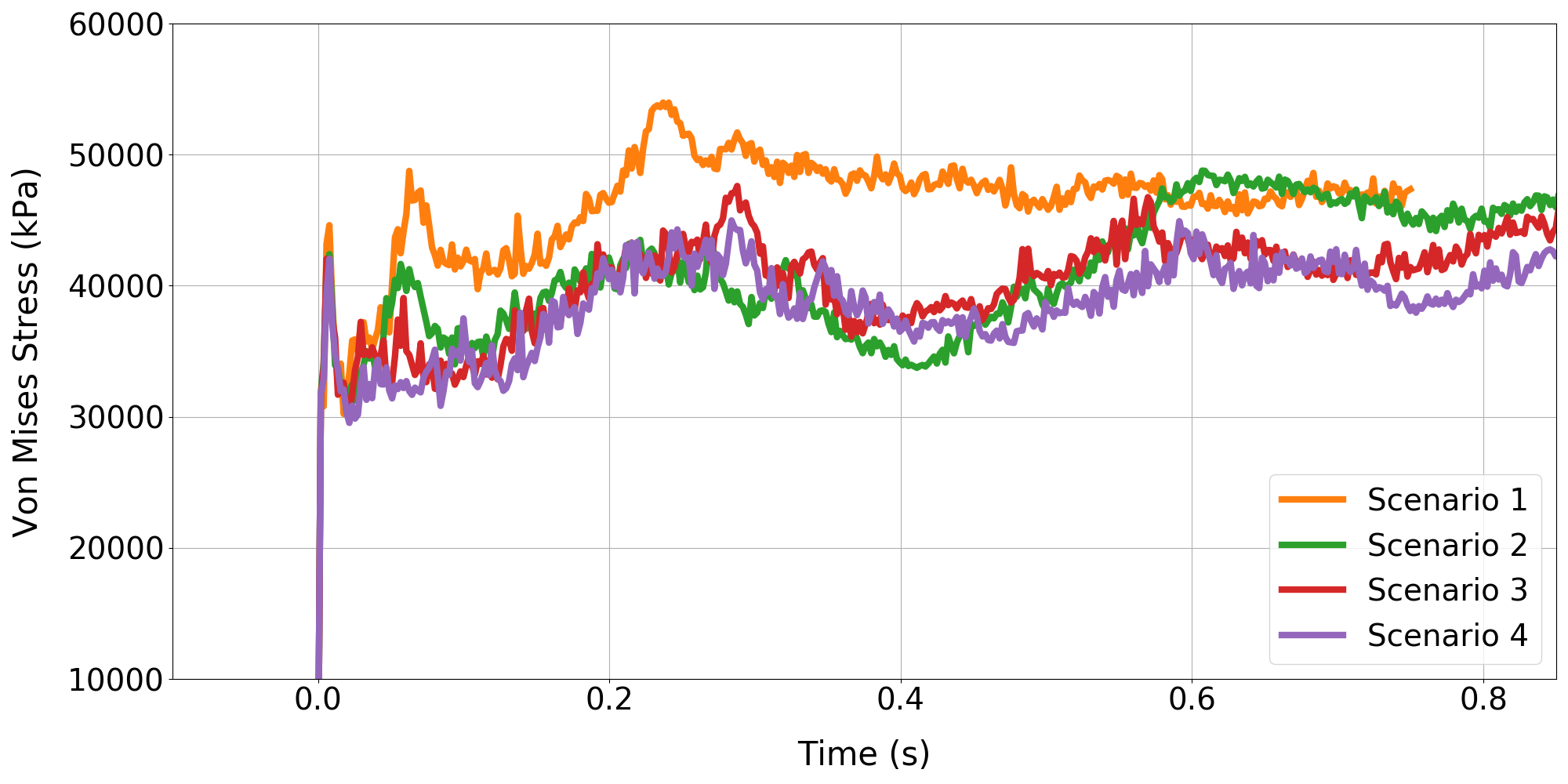}
\includegraphics[scale=0.30]{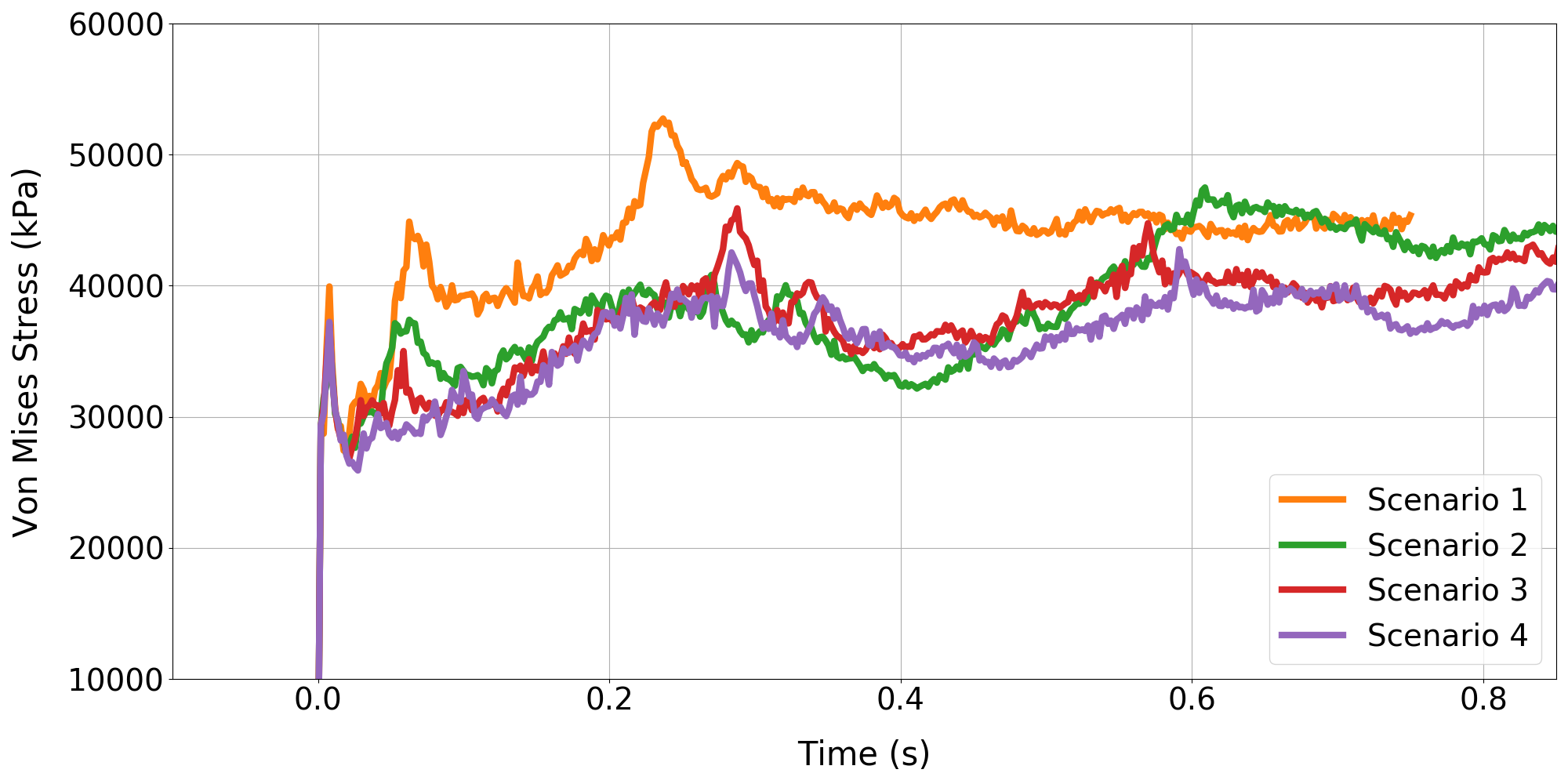}
\caption{Predicted time-histories of the von Mises stress, the maximum stress (top), the average of the
         maximum 80 stresses (bottom) during the parachute inflation.}
\label{fig:pid_vm}
\end{figure}

The time-histories of the maximum von Mises stress experienced by the parachute canopy during the simulated
inflation process discussed above are reported in \cref{fig:pid_vm}. Specifically, the maximum von Mises
stress and the average of the maximum 80 von Mises stress values, with the exceptions of several outliers
artificially caused by local contact self-penetrations, are reported for each case. For all cases, the
maximum von Mises stress is reached at or the same time as the peak drag force. Local peaks of this field
are obtained along with local peak drags when the disk gores are inflated, or in the partial collapse cycle.
The predicted maximum von Mises stress is about $5.0\times10^7$~Pa. This suggests that the parachute decelerator
system of Curiosity survived with a safety factor about 5. 

\section{Conclusions}
\label{sec:conclusion}
This paper presents a high-fidelity multi-physics Eulerian computational framework for Mars landing parachute
inflation simulations. A robust embedded boundary approach~(FIVER), equipped with efficient adaptive mesh
refinement and several novel techniques recently developed specifically for parachute inflation simulations,
including a homogenized porous wall model and a suspension line fluid-structure treatment, are discussed. 
The validation of the framework is presented by simulating the Mars atmosphere entry of the NASA Curiosity
Rover~\cite{cruz2014reconstruction}, i.e., a full-scale parachute inflation simulation in the low-density,
low-pressure Mars atmosphere. The canopy breathing, bow shocks and turbulent wake interactions are observed. 
The effects of fluid-suspension line interactions with the flow and bulk viscosity of the fluid medium, and
the sensitivity of the solution with respect to uncertain initial conditions are all highlighted. The predicted
and measured drag histories and the first peak forces reach reasonable agreements. To the best of our knowledge,
this is the first FSI simulation effort that matches the Mars landing data and suggests a new capability to provide
valuable insight for the future supersonic parachute design.

Although not accounted for in the present study, both the motion of the forebody and more sophisticated
constitutive modeling of the parachute canopy (incorporating for example multiscale and/or damage modeling)
are enabled by the proposed framework and will be considered in future simulations.

\section*{Acknowledgments}
Daniel Z. Huang, Philip Avery and Charbel Farhat acknowledge partial support by the Jet Propulsion Laboratory
(JPL) under Contract JPL-RSA No. 1590208, and partial support by the National Aeronautics and Space Administration
(NASA) under Early Stage Innovations (ESI) Grant NASA-NNX17AD02G. Parts of this work were completed at the
JPL, California Institute of Technology, under a contract with NASA.

\bibliographystyle{unsrt}
\bibliography{PID_SciTech}

\end{document}